\documentclass[twocolumn]{aastex631}
\usepackage{hyperref}
\usepackage{amsmath}
\newcommand{\RomanNumeralCaps}[1]
    {\MakeUppercase{\romannumeral #1}}

\accepted{April 16, 2024}

\begin{document}

\title{Detecting Population \RomanNumeralCaps{3} Stars through Tidal Disruption Events in the Era of JWST and Roman}

\correspondingauthor{Lixin Dai\\Rudrani Kar Chowdhury}
\email{lixindai@hku.hk\\ rudrani@hku.hk}

\author[0000-0003-2694-933X]{Rudrani Kar Chowdhury}
\affiliation{Department of Physics, The University of Hong Kong, Pokfulam Road, Hong Kong}

\author[0009-0004-2575-1924]{Janet N. Y. Chang}
\affiliation{Department of Physics, The University of Hong Kong, Pokfulam Road, Hong Kong}

\author[0000-0002-9589-5235]{Lixin Dai}
\affiliation{Department of Physics, The University of Hong Kong, Pokfulam Road, Hong Kong}

\author[0000-0002-5554-8896]{Priyamvada Natarajan}
\affiliation{Department of Astronomy, Yale University, New Haven, CT 06511, USA}
\affiliation{Department of Physics, Yale University, New Haven, CT 06520, USA}
\affiliation{Black Hole Initiative at Harvard University, 20 Garden St., Cambridge MA, 02138, USA}

\begin{abstract}
The first generation metal-free stars, referred to as population \RomanNumeralCaps{3} (Pop \RomanNumeralCaps{3}) stars, are believed to be the first objects to form out of the pristine gas in the very early Universe. Pop III stars have different structures from current generation of stars and are important for generating heavy elements and shaping subsequent star formation. However, it is very challenging to directly detect Pop \RomanNumeralCaps{3} stars given their high redshifts and short life-times. In this paper, we propose a novel method for detecting Pop \RomanNumeralCaps{3} stars through their tidal disruption events (TDEs) by massive black holes. We model the emission properties and calculate the expected rates for these unique TDEs in the early Universe at $z\sim10$. We find that Pop \RomanNumeralCaps{3} star TDEs have much higher mass fallback rates and longer evolution timescales compared to solar-type star TDEs in the local universe, which enhances the feasibility of their detection, although a good survey strategy will be needed for categorizing these sources as transients. We further demonstrate that a large fraction of the flare emissions are redshifted to infrared wavelengths, which can be detected by the James Webb Space Telescope and the Nancy Grace Roman Space Telescope. Last but not least, we find a promising Pop \RomanNumeralCaps{3} star TDE detection rate of up to a few tens per year  using the Nancy Grace Roman Space Telescope, based on our current understanding of the black hole mass function in the early Universe.
\end{abstract}


\section{Introduction} \label{sec:intro}

\noindent The first generation stars are believed to have formed from the pristine gas a few hundred million years after the Big Bang. These metal-free stars are commonly referred to as population \RomanNumeralCaps{3} (Pop \RomanNumeralCaps{3}) stars. Pop \RomanNumeralCaps{3} stars are believed to form in collapsed dark matter mini-halos with masses of around $10^6M_{\odot}$ at $z \sim 15-20$ \citep{Abel02, Bromm02, Yoshida06}. The initial mass function (IMF) of these short-lived Pop \RomanNumeralCaps{3} stars, is expected to be tilted to higher masses and hence they are likely more massive than the current generation of forming stars (Population I, or Pop I stars). While there is uncertainty at the present time regarding the Pop III IMF, their masses are believed to lie within the range of $30M_{\odot} - 300 M_{\odot}$. Pop III stars are expected to be key players in the early Universe and they are believed to be responsible for re-ionizing the inter galactic medium (IGM) \citep{Sokasian04, Johnson07, Kulkarni14}, and lifting the fog of the dark ages and jump-starting structure formation. Moreover, Pop \RomanNumeralCaps{3} stars produce metals in their core and therefore play a crucial role in bridging the gap between the metal-free primordial gas and higher-metallicity Population II (Pop II) and Pop I stars by polluting the IGM through supernovae explosions \citep{Chen22}. Hence, as metal polluters that shape subsequent generations of star formation, Pop III stars are an important population to detect directly as this will enable constraining their properties and better understand the evolution of subsequent generations of stellar populations. 

A plethora of numerical studies have been carried out to track the formation of Pop \RomanNumeralCaps{3} stars in the first few billion years of the Universe \citep{Abel02, Bromm09, Park21, saad22, Prole23}. In particular, recent studies show that Pop III stars do not form in isolation, and that typically multiple Pop \RomanNumeralCaps{3} stars can form inside a single minihalo \citep{clark11, Liu20, Park21b}. While it is generally believed that the Pop III star formation rate peaks around $z \sim 10$ \citep{Scannapieco03, Jaacks18}, there has been speculation that they could continue forming at lower redshifts if pristine gas pockets are available in galaxies \citep{Jimenez+2006,Liu20}.

Direct observation of Pop \RomanNumeralCaps{3} stars, however, turns out to be extremely challenging. So far only a few Pop \RomanNumeralCaps{3} like candidate objects have been identified \citep{vanzella20, welch22}: detected through serendipitous gravitational lensing \citep{Rydberg13, vikaeus22}; in pair-instability supernovae \citep{Hummel12, Reg20}; in gamma ray bursts \citep{desouza11, Mesler14, Lazar22} and recent claimed detection with JWST NIRSpec observations \citep{Wang22}. 

In this paper, we propose and explore in detail a novel detection channel for Pop \RomanNumeralCaps{3} stars motivated by the unfolding JWST observations and evidence for the possible existence of massive black holes (MBHs) formed via direct collapse of gas in place at these early epochs \citep{Bogdan+2023,Natarajan+2023}. Therefore, Pop \RomanNumeralCaps{3} stars stand to be revealed via the observation of their tidal disruption by MBHs in early galaxies. Recent JWST observations in combination with Chandra X-ray observations, have revealed and confirmed the presence of an accreting supermassive black hole (SMBH) in the source UHZ1 at $z \approx 10.1$, a mere 470 Myrs after the Big Bang \citep{Bogdan+2023} that is consistent with formation via direct collapse of gas \citep[e.g.,][]{Loeb94, Begelman06, choi13}.  

Tidal Disruption Events (TDEs) are produced whenever a star approaches a  massive black hole (MBH) and comes within the distance at which the tidal force of the black hole exceeds the self-gravity of the star. In such cases, the star will be destroyed by the tidal force, which in turn will produce a luminous tidal disruption event (TDE) \citep{Rees88, Evans89}. At the present time about 100 TDEs of Pop I stars in the local Universe have been observed \citep{Gezari21}, and these are typically associated with powerful emission observed in multiple wavebands including X-ray,  ultraviolet (UV) optical and radio \citep{Auchettl17, Alexander20, Saxton20, vv21}.

In this paper, we model the properties of Pop \RomanNumeralCaps{3} star TDEs. As these first stars are expected to exist primarily in high-redshift galaxies, their optical and UV emission will be redshifted rendering them detectable in infrared (IR) wavelengths. This study is extremely timely due to the recent launch and successful data stream from the James Webb Space Telescope \citep[JWST,][]{Gardner06} and planning underway for the soon to be launched Nancy Grace Roman Space Telescope \citep[Roman,][]{spergel15}. JWST and Roman operate in IR bands, covering the wavelength range of $600$ nm - $28000$ nm and $480$ nm -$2300$ nm respectively. With the high resolution Near Infrared Camera (NIRCam) and Near Infrared Spectrograph (NIRSpec) along with Mid-Infrared Instrument (MIRI), JWST has already started to revolutionize our understanding of the first galaxies and early MBHs \citep[e.g.,][]{Marco22, Hakim23, Hidenobu23, Bogdan+2023}. On the other hand, the planned extremely large field of view of Wide Field Instrument (WFI) on board Roman is designed to observe a vast area of the sky, making it possible to detect numerous faint objects simultaneously. One of the primary objectives of Roman is to shed light on the matter distribution over a large range of redshifts to understand the nature of dark matter and expansion history of the universe. Both missions are hence expected to yield a large sample of extremely faint high redshift sources.

Our paper is structured as follows: in Section \ref{sec:model}, we introduce the basic structure of Pop \RomanNumeralCaps{3} stars, based on which we calculate the key TDE parameters. We also introduce the fiducial model used to calculate the emission properties of these Pop \RomanNumeralCaps{3} TDEs. In Section \ref{sec:result}, we present our main results, predictions for the intrinsic and the observed emission spectra; luminosities and the light curves of Pop III star TDEs. We also calculate event rates and discuss the detection prospects for this new class of sources with JWST and Roman. We conclude with a summary of our key findings and implications of our results for unveiling the early Universe with these new probes in Section \ref{sec:discussion}.

\section{Modelling Pop III TDEs} \label{sec:model}

\subsection{Mass-Radius Relation of Pop \RomanNumeralCaps{3} Stars} \label{sec:starmodel} 

\noindent
We consider Pop \RomanNumeralCaps{3} stars in the main sequence stage. Massive Pop \RomanNumeralCaps{3} stars are primarily radiation-pressure dominated, so their structures can be approximated using a polytropic model with an index of $\gamma = 4/3$ or $n=\frac{1}{\gamma-1}=3$ \citep{Bromm01}. 
Throughout this work, we use the mass-radius relationship of Pop \RomanNumeralCaps{3} stars adopted from \citet{Bromm01}:
\begin{equation}\label{eq:M-R}
%    R_\star \simeq 10R_{\odot}\left(\frac{M_\star}{370M_{\odot}}\right)^{0.45}\left(\frac{Z}{10^{-9}}\right)^{0.09}
    R_\star \simeq 0.7R_{\odot}\left(\frac{M_\star}{M_{\odot}}\right)^{0.45}\left(\frac{Z}{10^{-9}}\right)^{0.09},
\end{equation}
where $M_\star$ is the stellar mass, $R_\star$ is the stellar radius, and $Z$ is the metallicity of Pop \RomanNumeralCaps{3} stars. When $Z$ increases, cooling through metal lines is enhanced  and the effective temperature of a Pop \RomanNumeralCaps{3} star drops, which means the average stellar density also decreases given $\rho_\star \propto T^3$ for a polytropic star with $n=3$ \citep{Fowler64}. Therefore, $R_\star$ increases with increasing $Z$. The value of the metallicity $Z$ considered has varied from zero to a very small non-zero number in previous treatments \citep{Bromm01, schaerer03, Murphy21, KG23}. Many studies show that there likely exists a critical metallicity at which the transition from metal-free Pop \RomanNumeralCaps{3} stars to metal-poor Pop \RomanNumeralCaps{2} stars occurs and that value is $\sim 10^{-3} - 10^{-5}Z_\odot$ depending on the IMF of Pop \RomanNumeralCaps{3} stars \citep{Bromm01b, Schneider02, yoshida04, wise12, Jaacks18}. Furthermore, \cite{Jaacks18} show that the mean gas metallicity increases with decreasing redshift $z$ and reaches $\sim 10^{-5}Z_{\odot}$ at $z \sim 10$. Hereafter, in this work in order to probe the metallicity dependence of our results, we calculate fiducial properties for Pop \RomanNumeralCaps{3} stars with two chosen values for the metallicity: $Z=10^{-5}$ and a lower metallicity of $Z=10^{-9}$.

\begin{figure}
    \includegraphics[width=8.5cm]{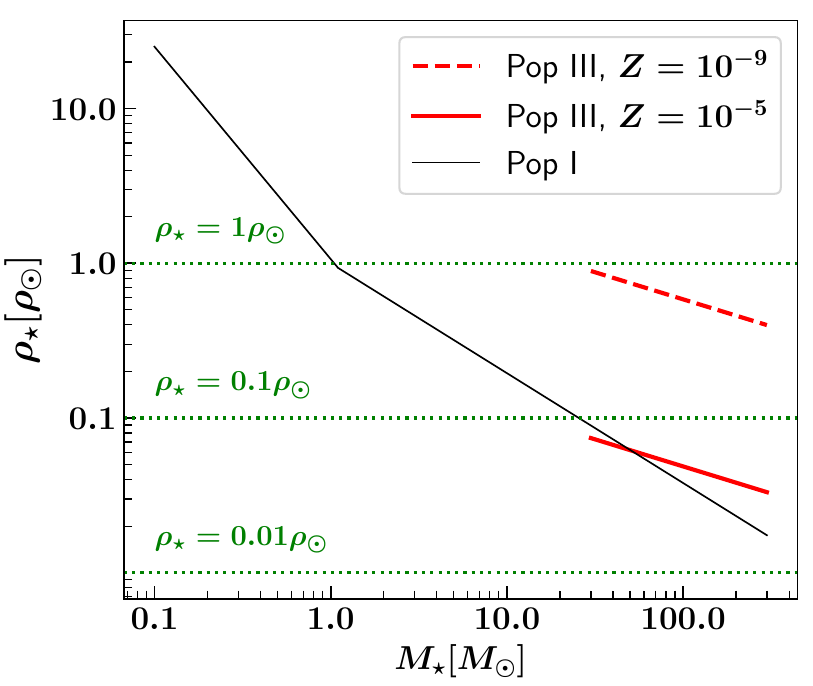}
    \caption{Average density ($\rho_\star$) of Pop \RomanNumeralCaps{3} stars with $Z=10^{-5}$ (red solid line) and $Z=10^{-9}$ (red dashed line) in the mass range of $30-300M_{\odot}$ compared to Pop I stars (black solid line) in the mass range of $0.1-300M_{\odot}$. All stars are in their main sequence stages. Pop \RomanNumeralCaps{3} stars typically have $\rho_\star$ $\sim0.01-1 \ \rho_{\odot}$ and those with lower metallicities or lower masses are denser. The stellar density of $1 \rho_{\odot}$ , $0.1 \rho_{\odot}$ and $0.01 \rho_{\odot}$ are marked  with green dotted lines.}
    \label{fig:Ms-rho}
\end{figure}

Given the stellar structure outlined above, we can calculate the the average stellar density $\rho_\star$ of Pop \RomanNumeralCaps{3} stars based on Eq. \ref{eq:M-R}. In Fig. \ref{fig:Ms-rho}, we show the variation of $\rho_\star$ for Pop \RomanNumeralCaps{3} stars as a function of their mass in the typical mass range of $30-300 M_{\odot}$ and compare that with the case of Pop I main-sequence stars. For the latter, we assume that their masses can extend to few hundreds of $M_{\odot}$ \citep{crowther10, Rickard23}, and we adopt the mass-radius relation from \citet{KW90}:
\begin{equation}\label{eq:M-R1}
    R_\star \sim
    \begin{cases}
      R_{\odot} \left(\frac{M_\star}{M_{\odot}}\right)^{0.8}; &  M_\star \leq 1M_{\odot} \\
      R_{\odot} \left(\frac{M_\star}{M_{\odot}}\right)^{0.57}; & M_\star > 1M_{\odot}
    \end{cases}
\end{equation}
It can be seen that for both types of stars, $\rho_\star$ drops as the $M_\star$ increases. Furthermore, Pop \RomanNumeralCaps{3} stars have $\rho_\star \sim \rho_{\odot}$ for the lower metallicity case $Z \sim 10^{-9}$ and $\rho_\star \sim0.1\rho_{\odot}$ as $Z$ approaches the upper limit of $10^{-5}$.

\subsection{Tidal Disruption of Pop \RomanNumeralCaps{3} Stars} \label{sec:TDE} 
\begin{figure}
    \includegraphics[width=8.5cm]{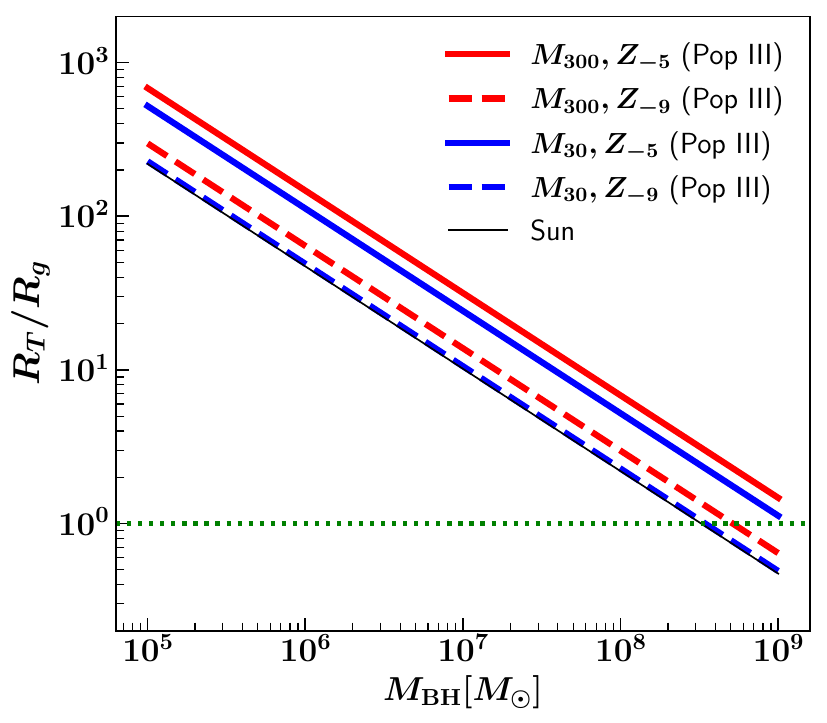}
    \caption{Tidal disruption radius $R_T$ (scaled by the BH gravitational radius $R_g$) of Pop \RomanNumeralCaps{3} stars and a solar-type star as a function of BH mass $M_{\rm BH}$. The red and blue solid or dashed lines represent Pop \RomanNumeralCaps{3} stars of mass $300M_\odot$ ($M_{300}$) and $30M_\odot$ ($M_{30}$) respectively with metallicity $Z=10^{-5}$ ($Z_{-5}$) or $10^{-9}$ ($Z_{-9}$). The black solid line shows a solar-type star for comparison. It can be seen that a Pop \RomanNumeralCaps{3} star, depending on its metallicity, has $R_T$ similar to or a few times larger than that of the Sun. The green dotted line indicates $R_T= R_g$, which gives the maximum mass of $M_{\rm BH}$ for a star to be disrupted inside the MBH event horizon.}
    \label{fig:Mbh-Rt}
\end{figure}

\noindent
A star is tidally disrupted by a MBH when it approaches within its tidal disruption radius ($R_T$). To first order, $R_T$ can be calculated using 
\begin{equation}\label{eq:M-Rt}
R_{ T} = R_\star\left(\frac{M_{\rm BH}}{M_\star}\right)^{1/3},
\end{equation}
where $M_{\rm BH}$ is the black hole (BH) mass. Therefore, for a fixed $M_{\rm BH}$, $R_T$ only depends on the average stellar density ($R_T \propto \rho_\star^{-1/3}$). Fig. \ref{fig:Mbh-Rt} shows the comparison of $R_T$ for Pop \RomanNumeralCaps{3} stars (of different masses and metallicities) and with that for solar type stars disrupted by MBHs with mass ranging from $M_{\rm BH}=10^5-10^9 M_{\odot}$. It can be seen that $R_T$ for Pop \RomanNumeralCaps{3} stars at very low metallicity ($Z\sim10^{-9}$) are similar to that of the Sun, while $R_T$ for Pop \RomanNumeralCaps{3} stars at relatively high metallicities ($Z\sim10^{-5}$) are a few times larger, which is consistent with the density comparison seen in Fig. \ref{fig:Ms-rho}. 

Moreover, since $R_T \propto M_{\rm BH}^{1/3}$ while the gravitational radius of a BH $R_g\equiv GM_{\rm BH}/c^2 \propto M_{\rm BH}$, there exists a upper limit of $M_{\rm BH}$ (called the Hill mass) beyond which an approaching star is swallowed by the MBH as a whole. One can see in Fig. \ref{fig:Mbh-Rt} that the high-metallicity ($Z\sim10^{-5}$) Pop \RomanNumeralCaps{3} stars can be disrupted by MBHs with masses up to $10^9 M_\odot$. This can be used as an important indicator for Pop \RomanNumeralCaps{3} TDE detection, since Pop I main-sequence stars can only be disrupted by MBHs with $M_{\rm BH}\lesssim 10^8 M_\odot$ except in the extreme case when the MBH has a close to maximal spin \citep{Kesden12}. Although the estimated mass of the BH in sources like UHZ1 (at $z \approx 10.1$) is estimated to be $\sim 4 \times 10^7 M_\odot$, it is expected that even more massive SMBHs could exist at these epochs although they are expected to be extremely rare. Meanwhile, MBHs with masses less than that of UHZ1 could be significantly more numerous at these early epochs.

\begin{figure*}
    \includegraphics[width=17.cm]{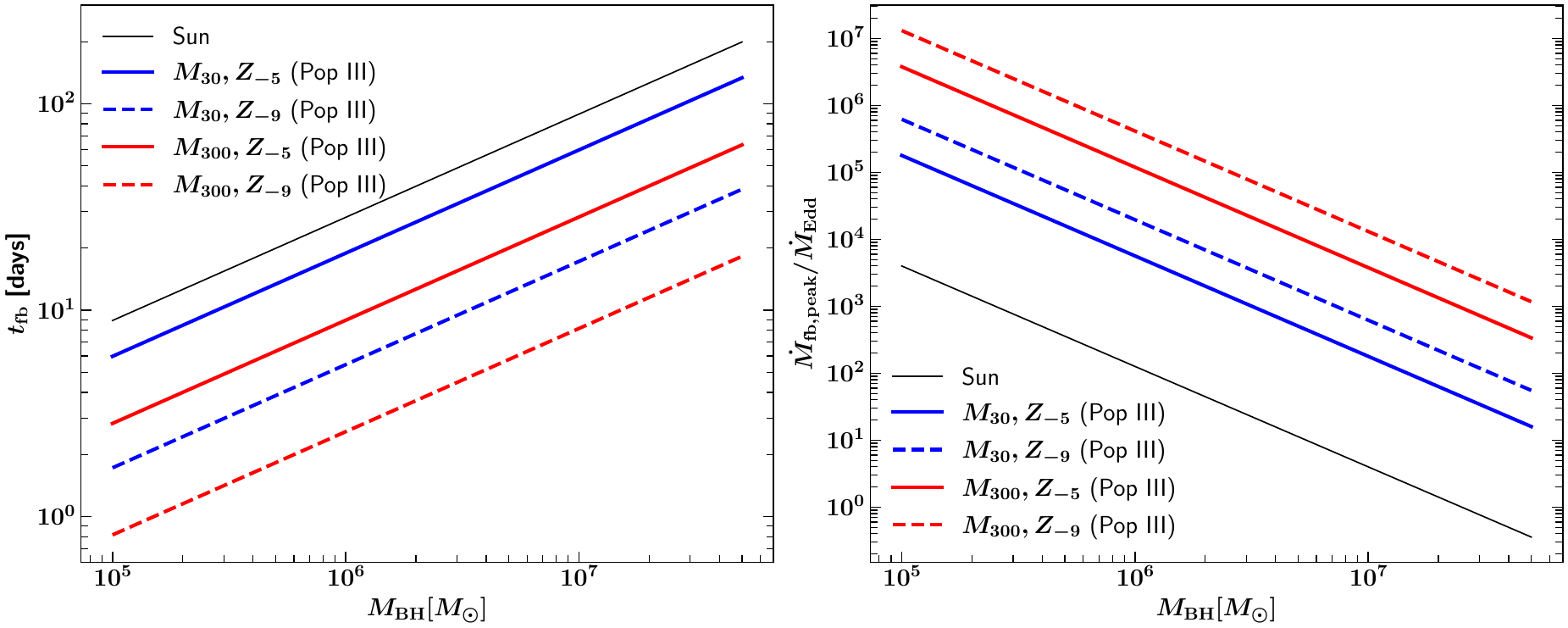}
    \caption{Left panel: Debris mass fallback time $t_{\rm fb}$ for tidal disruption of $300M_\odot$ ($M_{300}$, red) and $30M_\odot$ ($M_{30}$, blue) Pop \RomanNumeralCaps{3} stars with $Z=10^{-5}$ ($Z_{-5}$, solid lines) and $Z=10^{-9}$ ($Z_{-9}$, dashed lines) compared to a regular $1M_\odot$ star (black solid line). Right panel: Eddington ratio of the peak debris mass fallback rate $\dot{M}_{\rm fb, peak}$ for the same stars as in the left panel. Pop \RomanNumeralCaps{3} stars have shorter fallback times and much higher peak mass fallback rates compared to Pop I star TDEs.}
    \label{fig:tpeak}
\end{figure*}

After the star is disrupted, about half of the stellar debris is ejected from the system and the rest remains bound to the MBH with a spread in its specific binding energy. The rate that the bound stellar debris orbits back to the pericenter is called the debris mass fallback rate ($\dot M_{\rm fb}$), which can be calculated as \citep{Evans89, Phinney89}:
\begin{equation} \label{eq:Mfb}
    \dot{M}_{\rm fb} (t) \approx \frac{1}{3}\frac{M_\star}{t_{\rm fb}} \left ( \frac{t}{t_{\rm fb}} \right)^{-5/3},
\end{equation}
where the fallback time $t_{\rm fb}$ is the orbital time of the most tightly bound debris. 

In this work we adopt the results from  \citet{Guillochon13} (hereafter GR13) who performed high-resolution hydrodynamical simulations to study this disruption process. In particular, adopting polytropic stellar models, GR13 provide the following fitting formulae for the debris fallback time and the peak mass fallback rate:
\begin{equation}\label{eq:tfb}
%    t_{\rm fallback} = 0.2 B_{\gamma}M_6^{0.5}m_{*,30}^{-0.3}~\rm yr
    t_{\rm fb} = B_{\gamma}M_6^{1/2}m_\star^{-1}r_\star^{3/2}~\rm yr
\end{equation}

\begin{equation}
%    \dot{M}_{\rm fb, peak} = 150 A_{\gamma}M_6^{-0.5}m_{*,30}^{1.3}~M_{\odot}/\rm yr
    \dot{M}_{\rm fb, peak} = A_{\gamma}M_6^{-1/2}m_\star^2r_\star^{-3/2}~M_{\odot}\ \rm yr^{-1}
    \label{eq:Mfbpeak}
\end{equation}
where $M_6 = M_{\rm BH}/10^6M_{\odot}$ , $m_\star = M_\star/M_{\odot}$ and $r_\star = R_\star/R_{\odot}$. Furthermore, for a poly-tropic star with $\gamma = 4/3$, the coefficients are:
\begin{equation}
    B_{4/3} = \frac{-0.38670 + 0.57291\sqrt{\beta} - 0.31231\beta}{1-1.2744\sqrt{\beta} - 0.90053\beta},
\end{equation}
\begin{equation}
    A_{4/3} = \rm{exp}\left[\frac{27.261 - 27.516\beta + 3.8716\beta^2}{1-3.2605\beta -1.3865\beta^2}\right].
\end{equation}
Here $\beta\equiv R_T/R_p$ is called the penetration factor, where $R_p$ is the pericenter distance of the stellar orbit. A deep plunging orbit is commonly associated with $\beta \gg 1$, whereas mild or partial disruption is denoted by $\beta \sim 1$ \& $\beta \lesssim 1$ respectively. GR13 showed that stars with polytropic index $\gamma = 4/3$ are fully disrupted when $\beta \gtrsim 1.85$, and under this condition $\dot{M}_{\rm fb, peak}$ remains similar. Hence we mark this as the critical penetration factor $\beta_c$ and use $\beta=\beta_c=1.85$ for the calculations throughout this paper unless otherwise specified. For $\beta=1.85$, $B_{4/3}\sim0.08$, and $A_{4/3}\sim 3$.

For Pop \RomanNumeralCaps{3} stars, one can then use the mass-radius relationship (Eq. \ref{eq:M-R}) and rewrite Eq. \ref{eq:tfb} and \ref{eq:Mfbpeak} as:

\begin{equation}
%t_{\rm fb} = 0.6 B_{\gamma}M_6^{0.5}m_\star^{-0.3}z_\star^{0.1}~\rm yr
t_{\rm fb} \simeq 9 \left(\frac{B_{4/3}}{0.08}\right)\left(\frac{m_\star}{300}\right)^{-0.3}\left(\frac{Z}{10^{-5}}\right)^{0.1}M_6^{0.5}~\rm day
    \label{eq:tpeak1}
\end{equation}
\begin{equation}
\begin{split}
%    \dot{M}_{\rm fb, peak} = 1.7 A_{\gamma}M_6^{-0.5}m_\star^{1.3}z_\star^{-0.1}~M_{\odot}/\rm yr
    \dot{M}_{\rm fb, peak} \simeq 2.7 \times 10^3 \left(\frac{A_{4/3}}{3}\right)\left(\frac{m_\star}{300}\right)^{1.3}\left(\frac{Z}{10^{-5}}\right)^{-0.1}\\M_6^{-0.5}~M_{\odot}\ \rm yr^{-1}
    \label{eq:tpeak2}
\end{split}
\end{equation}

We plot $t_{\rm fb}$ and $\dot{M}_{\rm fb, peak}$ as functions of $M_{\rm BH}$ for various Pop \RomanNumeralCaps{3} stars compared to the Sun in Fig. \ref{fig:tpeak}. It can be seen that $t_{\rm fb}$ for Pop \RomanNumeralCaps{3} stars are usually around a few days, which is shorter compared to that of a solar-type star wherein $t_{\rm fb}$ $\sim$ a few tens of days for the case of $M_{\rm BH} =10^6M_\odot$. Furthermore, Pop \RomanNumeralCaps{3} stars have extremely high $\dot{M}_{\rm fb, peak}$ which exceeds that of Pop I star TDEs by several orders of magnitude. For example, when $M_{\rm BH}=10^6 M_\odot$, for Pop \RomanNumeralCaps{3} star TDEs $\dot{M}_{\rm fb, peak}$ can reach $10^{4-6} \dot{M}_{\rm Edd}$, where $\dot{M}_{\rm Edd} \equiv L_{\rm Edd}/\eta c^2$ is the Eddington accretion rate of the BH with $L_{\rm Edd}$ being the Eddington luminosity, $c$ the speed of light and $\eta$ is the radiative efficiency with a nominal value of 0.1. This hyper-Eddington debris mass fallback rate is a consequence of the very large masses of Pop \RomanNumeralCaps{3} stars coupled with their short debris mass fallback timescales.

After the peak of the flare, the debris mass fallback rate drops with time following $t^{-5/3}$. At late times, the fallback rate should transit from super-Eddington to sub-Eddington. We can calculate the timescale over which the fallback rate stays super-Eddington using $\dot{M}_{\rm fb}=\dot{M}_{\rm Edd}$:

\begin{equation}
%    t_{\rm Edd} \approx 0.5 \left(\frac{m_\star}{M_6}\right)^{3/5}\left(\frac{t_{\rm fb}}{\rm day}\right)^{2/5} \rm yr
    t_{\rm Edd} \approx 38.5 \left(\frac{m_\star}{300}\right)^{3/5}\left(\frac{t_{\rm fb}}{10~\rm day}\right)^{2/5}M_6^{-3/5} \ \rm yr
\end{equation}

\begin{figure}
    \includegraphics[width=8.5cm]{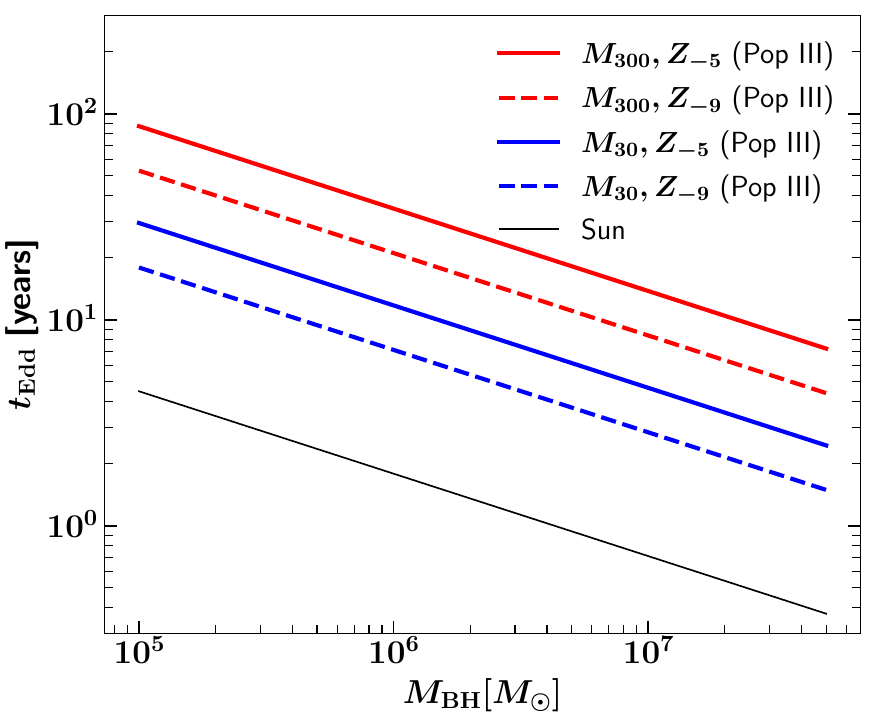}
    \caption{The Eddington timescale defined as the timescale that the debris mass fallback rate stays above the Eddington accretion rate. It is seen that Pop \RomanNumeralCaps{3} star  TDEs can stay super-Eddington for very long timescales (up to hundreds of years).}
    \label{fig:tedd}
\end{figure}

This timescale is shown in Fig. \ref{fig:tedd}. It is clearly seen that Pop \RomanNumeralCaps{3} TDEs have super-Eddington fallback rates for a much longer duration (up to hundreds of years) compared to a solar type star TDE (a few years) considering $M_{\rm BH} = 10^6M_\odot$.

We also note that there exist alternative models for the disruption process of stars. For example, a recent semi-analytical work by \citet{B24} showed that the fallback time could be almost independent of the stellar mass (in Pop I star TDEs). Yet these different models likely lead to consistent first-order results for the peak fallback rates which matters the most for the observed flux (see Appendix \ref{app:comparison} for comparison between models and more discussion).

\subsection{TDE Optical/UV Emission Model} \label{sec:wind}

\noindent
The extremely high debris mass fallback rate seen in Fig. \ref{fig:tpeak} means that even if just a small fraction of the debris reaches the vicinity of the MBH, a super-Eddington accretion flow could result, which will launch powerful winds due to the large radiation pressure \citep{Dai18, defu23}. Furthermore, outflows are also expected to be powered by debris stream collisions \citep{Shiokawa15, Bonnerot19, Lu20}. 

It has been proposed that these outflows are responsible for producing the optical/UV emission observed from TDEs in the local universe \citep{Loeb97, Strubbe09, Lodato11, Roth16, Metzger16, Roth20, Dai21}. There have also been sophisticated numerical simulations studying super-Eddington BH accretion, outflow and emission \citep{Sadowski16, Dai18, Jiang19, Thomsen22}. However, no simulation has been done yet to study super-Eddington accretion flows with Eddington ratios as high as $10^{5-7}$. Given lack of guidance from simulations, we adopt an analytical model proposed by \citet[][hereafter SQ09]{Strubbe09} to calculate the emission properties of Pop III star TDEs, while making a few changes to adapt model parameters consistent with more recent studies.

\begin{figure}
    \includegraphics[width=8.5cm]{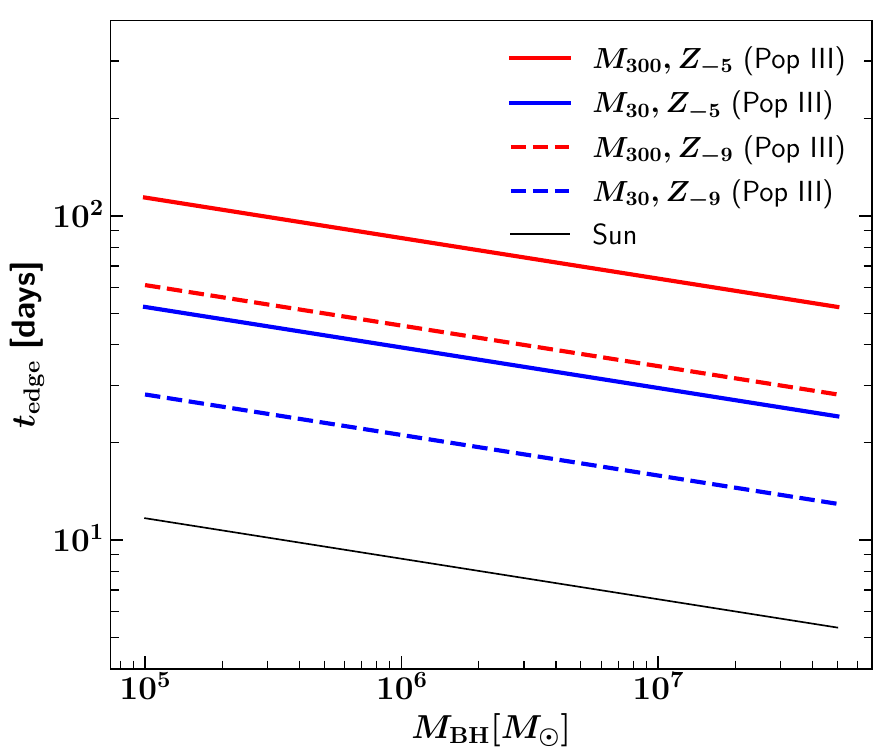}
    \caption{The time that the photo-sphere transits from expansion to recession  ($t_{\rm edge}$)  as a function of $M_{\rm BH}$ for various types of stars. Colour scheme is the same as Fig. \ref{fig:tpeak}. $t_{\rm edge}$ is longer for the Pop \RomanNumeralCaps{3} stars compared to solar-type stars.}
    \label{fig:t_edge}
\end{figure}

We first provide a brief summary of the SQ09 model below. SQ09 assume that during the early, super-Eddington phase a sizable fraction of the fallback material constitutes an outflow. Furthermore, they also assume that a fraction of the wind kinetic energy is converted to thermal energy, which in turn produces black-body radiation. This assumption appears to hold even when the outflows are powered by debris stream collisions, and it is supported by recent simulations \citep{Dai18, Zanazzi20, Andalman22, Thomsen22}. SQ09 assume a spherical wind geometry and a constant wind velocity, under these circumstances, the wind density profile can be approximated as:
\begin{equation}
    \rho (r)=
    \begin{cases}
        \dot{M}_{\rm wind}/(4 \pi r^2v_{\rm wind}), &  r \lesssim R_{\rm edge} \\
      0, & \text{outside}
    \end{cases}
\end{equation}
Here the wind mass rate $\dot{M}_{\rm wind}$ is assumed to be a constant fraction of the debris mass fallback rate so that $\dot{M}_{\rm out} \equiv f_{\rm out}\dot{M}_{\rm fb}$. We use a fiducial value $f_{\rm out}=0.5$ (instead of $f_{\rm out}=0.1$ used by SQ09), inspired by recent numerical simulations of super-Eddington accretion flows \citep[e.g.][]{Dai18, Jiang19, Thomsen22}. We recognize that the exact value of $f_{\rm out}$ has a dependence on the Eddington ratio, but our results do not significantly vary when $f_{\rm out}>0.5$. $R_{\rm edge} \equiv v_{\rm wind}t$ denotes the edge of the wind. The wind velocity $v_{\rm wind}$ is assumed to be the escape velocity $v_{\rm esc,L}\equiv \sqrt{2GM_{\rm BH}/R_L}$ at the wind-launching radius $R_L$, which is set to be $2 R_p$ (the circularization radius). Therefore, we have $v_{\rm esc,L} = \sqrt{GM_{\rm BH}/R_p}$, which can be further simplified as:

\begin{equation}
    v_{\rm esc,L} = 0.1 \beta^{1/2}m_\star^{1/6}r_\star^{-1/2}M_6^{1/3}c
    %v_{\rm esc,L} = 0.1 \beta^{0.5}m_\star^{0.2}r_\star^{-0.5}M_6^{0.3}c 
\end{equation}

Furthermore, SQ09 calculate the temperature of the wind by assuming that the gas thermal energy density and kinetic energy density are similar at the wind-launching site:
\begin{equation}
    aT_L^4 = \frac{1}{2}\rho_{\rm fb,L} v_{\rm esc,L}^2.
\end{equation}
Here $a=\sigma/4$ with $\sigma$ being the Stefan-Boltzmann constant, $T_L$ is the gas temperature at $R_L$, and $\rho_{\rm fb,L}$ is the gas density at $R_L$ which can be calculated as $\rho_{\rm fb,L} \equiv \dot{M}_{\rm fb}/(4 \pi R_L^2v_{\rm esc,L})$. It is further assumed that the wind expands adiabatically so that its temperature scales with gas density: 
\begin{equation} \label{eq:adiabatic}
    T \propto \rho^{1/3}.
\end{equation}

The photosphere radius of the wind, $R_{\rm ph}$, is located where 
\begin{equation}\label{eq:radius}
    R_{\rm ph}\kappa_s \rho (R_{\rm ph}) \sim 1,
\end{equation} 
where $\kappa_s$ is the electron scattering opacity which is taken to be $0.35 \ \rm cm^2 g^{-1}$ assuming the Hydrogen abundance for Pop \RomanNumeralCaps{3} stars is $\sim 75\%$, while rest of the mass is mostly dominated by Helium \citep{Bromm09}. 

Initially, the wind gas density is very high due to the large fallback rate, so $R_{\rm edge} \kappa_s \rho (R_{\rm edge})\gg 1$ and the photo-sphere almost coincides with the edge of the wind, which gives :
\begin{equation} \label{eq:Rin}
%    R_{\rm ph} \approx 5.2 \times 10^{14} f_v \beta_{1.8}^{0.5}m_{*,30}^{-0.03}M_6^{0.3} \left(\frac{t}{\rm {day}}\right)~\rm cm
%    R_{\rm ph} \approx R_{\rm edge} = 1.3 \times 10^{3} f_v \beta^{0.5}m_\star^{0.2}r_\star^{-0.5}M_6^{-0.7} \left(\frac{t}{\rm {day}}\right)~ R_s
    R_{\rm ph} \approx R_{\rm edge} = 1.3 \times 10^{3} f_v \beta^{1/2}m_\star^{1/6}r_\star^{-1/2}M_6^{-2/3} \left(\frac{t}{\rm {day}}\right)~ R_s
\end{equation}
where $R_s$ is the Schwarzschild radius of a non-spinning black hole. The effective temperature of the photo-sphere at this stage is given by:

\begin{equation}\label{eq:Tin}
\begin{split}
T_{\rm ph} \approx 1.5 \times 10^4 f_v^{-1/3}m_\star^{-7/72}r_\star^{1/24} M_6^{5/36}\\
\left(\frac{t}{\rm day}\right)^{-7/36}\left(\frac{t_{\rm fb}}{\rm yr}\right)^{-1/18} \rm K 
\end{split}
\end{equation}

As $R_{\rm edge} \kappa_s \rho (R_{\rm edge})\propto \dot{M}_{\rm wind}/R_{\rm edge}$ decreases with $t$, after a certain time, $R_{\rm edge} \kappa_s \rho (R_{\rm edge})$ drops to 1 and the photo-sphere starts to recede afterwards.
This transition time is calculated to be:
\begin{equation} \label{eq:t_edge}
    t_{\rm edge} \approx 9.6 f_{\rm out,0.5}^{3/8}f_v^{-3/4}B_\gamma^{1/4}M_6^{1/8}R_{ p,3R_s}^{3/8}m_\star^{1/8}r_\star^{3/8}~\rm day
\end{equation}
where $f_{\rm out,0.5} \equiv f_{\rm out}/0.5$, and $f_v = v_{\rm wind}/v_{\rm esc,L}$ (this fraction is set as $1$ throughout this paper unless otherwise specified). As we will show later, $t_{\rm edge}$ is also the time when the flare bolometric luminosity reaches the peak. We plot $t_{\rm edge}$ for Pop \RomanNumeralCaps{3} stars and the Sun in Fig. \ref{fig:t_edge}. It is clearly seen that the transition occurs on timescales of a few tens of days for Pop \RomanNumeralCaps{3} stars as compared to a few days for the Sun in case of $M_{\rm BH}=10^6M_\odot$. This leads to larger photo-sphere radii and higher peak luminosities for Pop \RomanNumeralCaps{3} TDEs.

After the photo-sphere starts to recede, the photo-sphere radius and temperature at this later stage can be estimated following Eq. \ref{eq:adiabatic} and Eq. \ref{eq:radius} and given as:
\begin{align}\label{eq:Rf}
\begin{split}
 R_{\rm ph} &= \kappa_s\dot{M}_{\rm out}/(4\pi v_{\rm wind}) \\
 &\approx 5.4 f_{\rm out,0.5}f_v^{-1} \left(\frac{\dot{M}_{\rm fb}}{\dot{M}_{\rm Edd}}\right)R_{p,3R_s}^{1/2}R_s
\end{split}
\end{align}

\begin{equation}\label{eq:Tf}
T_{\rm ph} \approx 2.5 \times 10^5 f_{\rm out,0.5}^{-1/3} f_v^{1/3} \left(\frac{\dot{M}_{\rm fb}}{\dot{M}_{\rm Edd}}\right)^{-5/12}M_6^{-1/4}R_{p,3R_s}^{-7/24}~ \rm K
%T_{\rm ph} \approx 2.5 \times 10^5 f_{\rm out,0.5}^{-0.3} f_v^{0.3} \left(\frac{\dot{M}_{\rm fallback}}{\dot{M}_{\rm Edd}}\right)^{-0.4}M_6^{-0.2}R_{p,3R_s}^{-0.3}~ \rm K
\end{equation}
where $R_{\rm p,3R_s} = R_p/3R_s$.

Specifically for Pop \RomanNumeralCaps{3} stars, using their mass-radius relationship (Eq. \ref{eq:M-R}), we have:

\begin{equation}
\begin{split}
%    t_{\rm edge} \approx 8.4 f_{\rm out,0.5}^{3/8}f_v^{-3/4}B_\gamma^{1/4}m_\star^{13/44}M_6^{1/8}R_{p,3R_s}^{3/8}z_\star^{3/88}
    t_{\rm edge} \simeq 85 f_v^{-3/4}\left(\frac{f_{\rm out}}{0.5}\right)^{3/8}\left(\frac{B_\gamma}{0.08}\right)^{1/4}\left(\frac{\beta}{1.85}\right)^{-3/8}\\\left(\frac{m_\star}{300}\right)^{15/44}\left(\frac{Z}{10^{-5}}\right)^{3/44}M_6^{-1/8} \ \rm day
    \end{split}
\end{equation}

Before  $t_{\rm edge}$:
\begin{equation}
    R_{\rm ph} \propto M_6^{1/3}m_\star^{-2/33}z_\star^{-1/22},
\end{equation}
\begin{equation}
    T_{\rm ph} \propto M_6^{1/9}m_\star^{-2/33}z_\star^{-1/264}.
\end{equation}
After $t_{\rm edge}$:
\begin{align}
    R_{\rm ph} \propto m_\star^{28/33}z_\star^{3/22},\\
    T_{\rm ph} \propto M_6^{2/9} m_\star^{-4/11}z_\star^{-17/264}.
\end{align}

\begin{figure*}
\begin{center}
     \includegraphics[width=18.1cm]{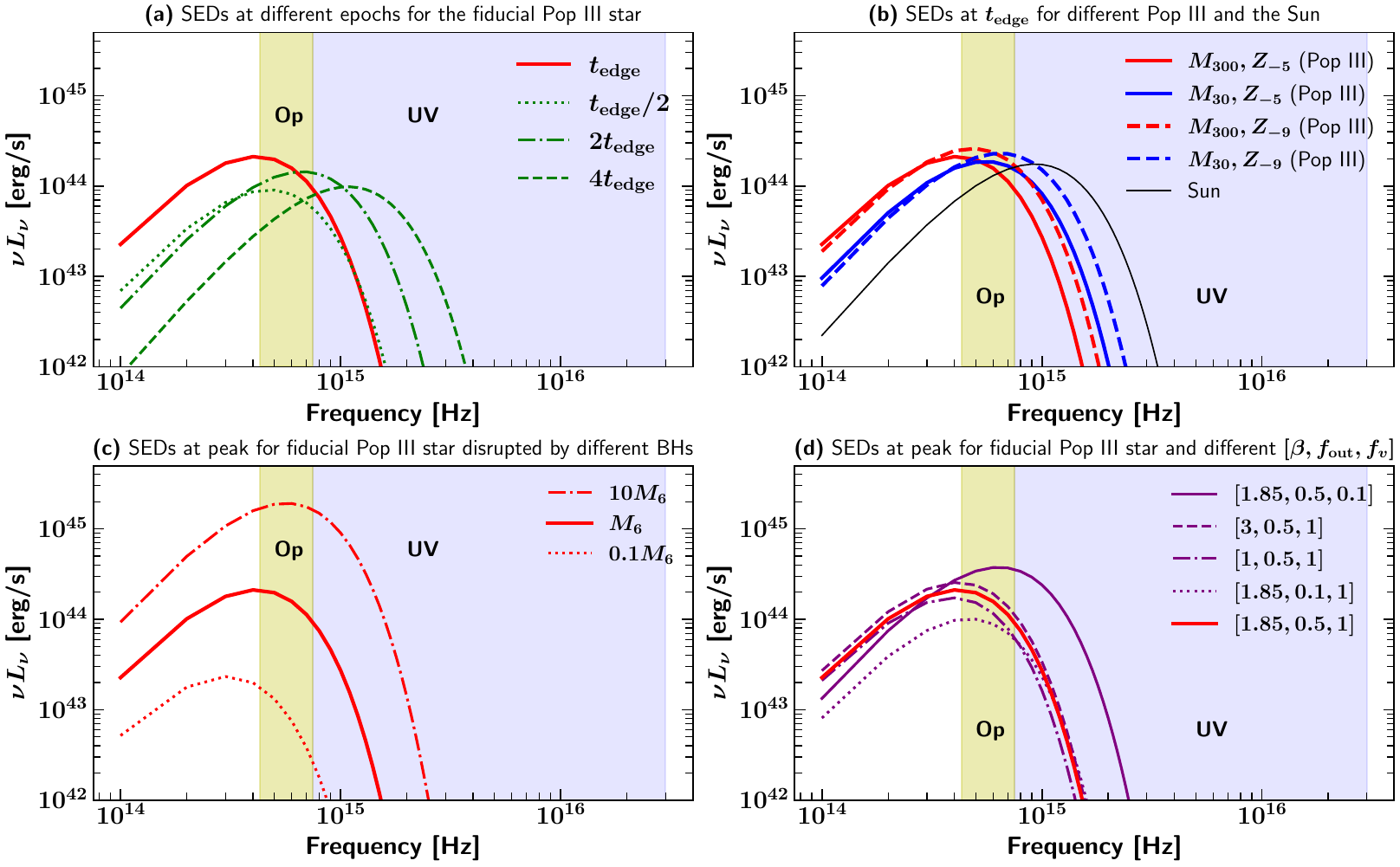}
    \caption{SEDs of the Pop \RomanNumeralCaps{3} star tidal flares in the rest frame of the host galaxy. The default parameter set used for plotting this figure and the subsequent ones is: $M_{\rm BH}=10^6M_\odot,~M_\star=300M_\odot~(M_{300}),~Z=10^{-5}~(Z_{-5}),~f_{\rm out} = 0.5,~f_v=1$, and $\beta = \beta_c = 1.85$  (the red solid curve) unless specified otherwise.
    \textbf{(a)} The evolution of the SED at different epochs: $t_{\rm edge}/2$, $t_{\rm edge}$ (peak luminosity), $2 \times t_{\rm edge}$ and $4 \times t_{\rm edge}$. \textbf{(b)} The SEDs at  $t= t_{\rm edge}$ for various Pop \RomanNumeralCaps{3} stars of different masses or metallicities and a solar-type star. \textbf{(c)} The dependence of the SED (at  $t= t_{\rm edge}$) on $M_{\rm BH}$.  \textbf{(d)} The dependence of the SED (at  $t= t_{\rm edge}$) on other parameters including the  mass outflow fraction ($f_{\rm out}$), the ratio between wind velocity and escape velocity ($f_v$) and the stellar orbital penetration parameter $\beta$ (with $\beta_c=1.85$ being the critical value for full tidal disruption).}
    \label{fig:SED}
\end{center}
\end{figure*}

We note that this first-order calculation in SQ09 assumes that the outflows start to be launched around the peak of the debris mass fallback rate and they ignore the outflows produced prior. The luminosity increases in the initial phase as the wind builds up and therefore the characteristic rising time for the flare will be $t_{\rm edge}$ instead of $t_{\rm fb}$. While this assumption can cause issues for a solar-type star TDE which has $t_{\rm edge}\lesssim t_{\rm fb}$, it works well for Pop \RomanNumeralCaps{3} star TDEs in which $t_{\rm edge}\gg t_{\rm fb}$. Therefore, Pop \RomanNumeralCaps{3} star TDE flares should rise on timescales of  $t_{\rm edge}$, i.e., a few weeks to months in the intrinsic TDE frame.

Furthermore, SQ09 also consider disk emission after the fallback rate drops to the sub-Eddington level. At this phase, the gas can radiate efficiently and a thin accretion disk is expected to form, which produces emission mainly in the X-ray and EUV bands. However, for Pop \RomanNumeralCaps{3} star TDEs, as seen from Fig. \ref{fig:tedd}, the fallback rate stays super-Eddington for a very long time (tens to hundreds of years). Therefore, in this work, we focus only on the emission produced from the winds launched in the super-Eddington phase and ignore the disk emission at very late times. 

\section{Results} \label{sec:result}

\subsection{Pop \RomanNumeralCaps{3} Star TDE Intrinsic Emission Spectrum and Luminosity}

\noindent
We first calculate the TDE luminosity and spectral energy distribution (SED) in the rest frame of the host galaxy based on the properties of the photo-sphere described in Section \ref{sec:wind}. 
Assuming blackbody emission, the energy spectrum and bolometric luminosity can be computed using: 
\begin{equation}\label{eq:sed}
    \nu L_{\nu} = 4 \pi^2 R_{\rm ph}^2 \nu B_{\nu} (T_{\rm ph}),
\end{equation}
\begin{equation}\label{eq:bol}
    L = 4 \pi \sigma R_{\rm ph}^2T_{\rm ph}^4.
\end{equation}
We plot the emission SEDs in Fig. \ref{fig:SED} and explore the dependence on various model parameters. The default set of parameters for a fiducial Pop III star TDE model is: $M_{\rm BH}=10^6M_\odot,~M_\star=300M_\odot,~Z=10^{-5},~f_{\rm out} = 0.5,~f_v=1$ and $\beta = \beta_c = 1.85$. We stick to these parameters for calculations unless specified otherwise.

\begin{figure*}
\begin{center}
     \includegraphics[width=18.cm]{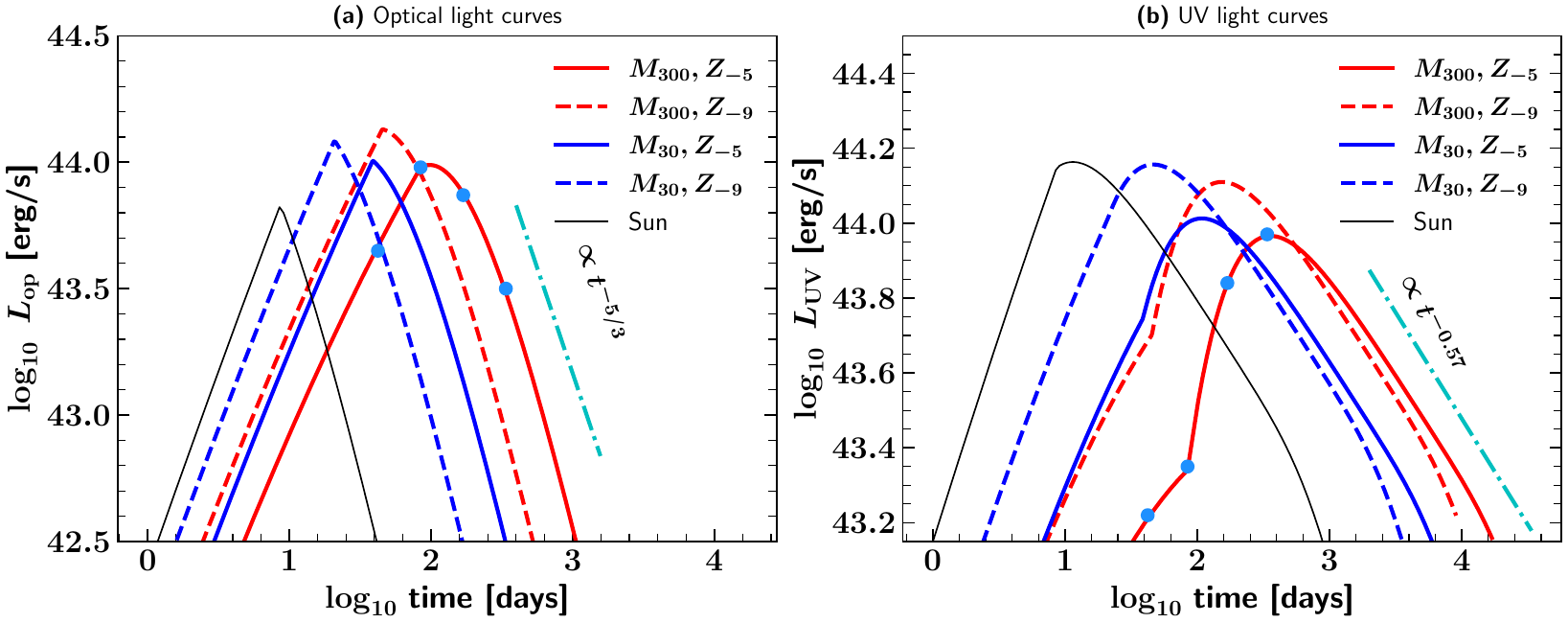}
    \caption{Optical (left) and UV (right) light curves of the Pop \RomanNumeralCaps{3} star TDEs around a MBH with $M_{\rm BH}= 10^6M_\odot$ in comparison to that that of a solar-type star. The color and line styles for different stars are the same as in Fig. \ref{fig:tpeak}. For the fiducial Pop \RomanNumeralCaps{3} star TDE model (the red solid curve), we  mark the four epochs used in Fig. \ref{fig:SED} (a) (i.e., $t_{\rm edge}/2$, $t_{\rm edge}$, $2 \times t_{\rm edge}$ and $4 \times t_{\rm edge}$) using light blue dots. The late-time light curves are fitted with power-law functions that are plotted using cyan, dot-dashed lines. 
    We see that the Pop \RomanNumeralCaps{3} star TDEs have much longer rise and decay timescales compared to standard TDEs of normal stars. Furthermore, for a Pop \RomanNumeralCaps{3} star TDE, while its optical light curves follow the debris fallback rate relatively closely and peaks around $t_{\rm edge}$, its UV light curve peaks at a later time and decays much more slowly due to the temperature evolution of the flare.}
    \label{fig:lc}
\end{center}
\end{figure*}

We illustrate the evolution of the emission from a single TDE in Fig. \ref{fig:SED}(a), which shows the SEDs at different epochs for the fiducial case. It is seen that initially the luminosity increases and the peak of the SED shifts slightly towards lower frequency, as a result of increasing $R_{\rm ph}$ and decreasing $T_{\rm ph}$ during this phase. After $t_{\rm edge}$,  $R_{\rm ph}$ recedes while $T_{\rm ph}$ increases, so the SED evolves in reverse, i.e., the luminosity decreases and the SED peak shifts towards higher frequency. 

Fig. \ref{fig:SED}(b) shows the SEDs at $t=t_{\rm edge}$ (the peak of the flare) from the TDEs of Pop \RomanNumeralCaps{3} stars of different masses and metallicities in comparison to that of a solar type star, all around a $10^6M_\odot$ MBH. It can be immediately noticed that a significant fraction of the emission energy resides in the UV/optical wavelength regimes for all cases. The SEDs of Pop \RomanNumeralCaps{3} star TDEs, compared to that of the solar-type TDE, have lower peak temperatures (with the SED peaking in the optical band instead of UV) and slightly larger luminosities. However, the mass or metallicity of a Pop \RomanNumeralCaps{3} star does not make a significant difference to its tidal flare emission. Increasing the metallicity leads to a slightly smaller luminosity and increasing the stellar mass shifts the SED towards slightly longer wavelengths.

Fig. \ref{fig:SED}(c) shows the impact of $M_{\rm BH}$ on the SED. It can be seen that more massive BHs produce more luminous flares. Interestingly, the  recent detection of sources such as UHZ1 \citep{Bogdan+2023} and GNz-11 \citep{Maiolino+2024} suggest that massive black holes of $M_{\rm BH}\gtrsim 10^{6-7} M_\odot$ at $z\geq10$ could be much more abundant than previously estimated. Therefore, it is likely that some Pop III TDEs can produce very luminous flares with intrinsic bolometric luminosity larger than $10^{45}\ \rm{erg \ s^{-1}}$.

Finally, Fig. \ref{fig:SED}(d) shows how various other parameters, namely, $f_{\rm out}$ (outflow fraction), $f_v$ (ratio between wind velocity and escape velocity) and $\beta$ (stellar orbital penetration parameter), affect the SED.  One can see that overall the choice of these parameters only mildly impacts the SED. We note that increasing $f_{\rm out}$ or decreasing $f_v$ generally both lead to a larger peak luminosity. A lower $\beta$ means the star is only partially disrupted, which will reduce the outflow mass, so the effect induced is similar to having a lower $f_{\rm out}$. Increasing $\beta$ beyond $\beta_c$ only slightly increases the peak fallback rate so it barely also affects the peak luminosity of the flare.

Next we consider the evolution of the flare flux in specific wavelength bands. Fig. \ref{fig:lc} shows the optical ($430 - 750$ THz) and UV ($750 - 3 \times 10^{4}$ THz) light curves of TDEs of Pop \RomanNumeralCaps{3} and solar-type stars by a MBH of  $M_{\rm BH}=10^6M_\odot$. For the fiducial Pop \RomanNumeralCaps{3} star model, we also mark the epochs ($t=t_{\rm edge}/2,t_{\rm edge}, 2t_{\rm edge},4t_{\rm edge}$) on the light curves. One can see that both the optical and UV luminosities increase in the initial phase when $R_{\rm ph}$ traces the edge of the wind. However, the behavior of the UV and optical light curves after $t_{\rm edge}$ are different.
For a solar-type star, since the peak of the flare SED stays in the UV regime, both the UV and the optical light-curve reach the peak around $t_{\rm edge}$ and decay afterwards. However, for a Pop \RomanNumeralCaps{3} star, the TDE flare SED shifts from UV to optical bands around $t_{\rm edge}$ and shifts back to UV band afterwards. Therefore, while the tidal flare optical light curve still peaks at $t_{\rm edge}$, the UV light curve continues to rise for tens to few hundreds of days after $t_{\rm edge}$. Furthermore, at late times, while the optical luminosity decays rather closely following the debris mass fallback rate ($\propto t^{-5/3}$), the UV light curve has a much shallower slope ($\propto t^{-0.57}$) due to the temperature evolution in this phase. This also means that the UV light curve has an evolution timescale longer than that of the optical light curve.

In summary, the bolometric luminosity of a Pop \RomanNumeralCaps{3} star TDE in the rest frame of the host galaxy increases with a larger $M_{\rm BH}$; a smaller metallicity; a higher outflow fraction, or a slower wind velocity, which all likely promote the potential observability of these TDEs. Furthermore, a higher Pop III mass will additionally serve to lengthen the evolution timescale of these light curves. We calculate the properties of the observed, and redshifted TDE emission in the next section.

\subsection{Pop \RomanNumeralCaps{3} Star TDEs: Observed Fluxes and Light Curves} \label{sec:jwst}

\begin{figure*}
\begin{center}
     \includegraphics[width=18.5cm]{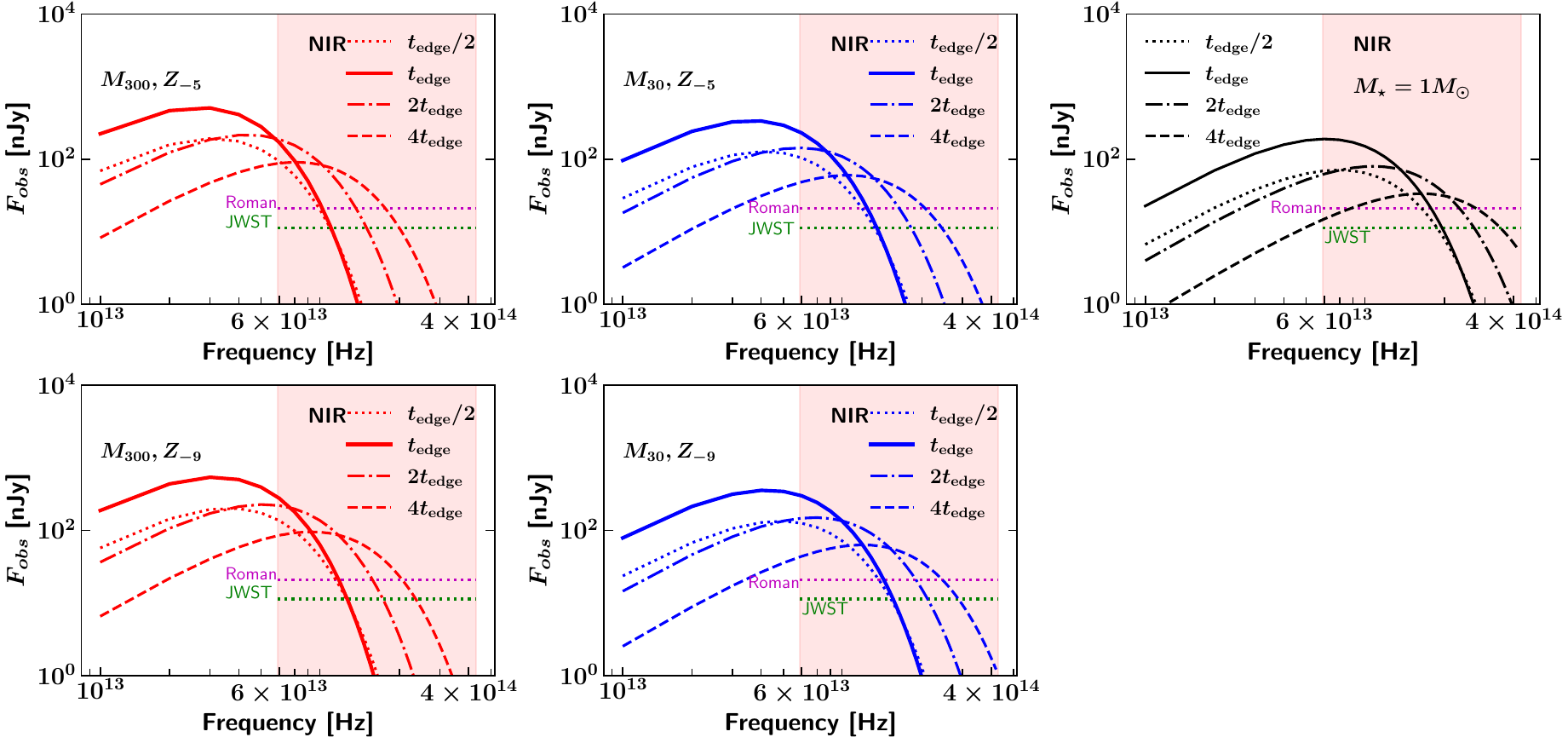}
    \caption{The observed fluxes of TDEs of Pop \RomanNumeralCaps{3} stars of mass $300M_\odot$ ($M_{300}$, left panel) and $30M_\odot$ ($M_{30}$, middle panel) with different metallicities of $10^{-5}$ ($Z_{-5}$, top panel) and $10^{-9}$ ($Z_{-9}$, bottom panel) compared to a solar type star (right panel) at different epochs ($t_{\rm edge}/2,t_{\rm edge},2t_{\rm edge},4t_{\rm edge}$). Green dotted line shows the sensitivity limit of F150W filter of NIRCam. Magenta dotted line represents the sensitivity limit of F106 filter of WFI, which has the best sensitivity limit of all the four filters (F106, F129, F158, F184) used in the High Latitude Wide Area Survey. Red shaded regions indicate JWST NIRcam bands. It can be seen that a large part of the fluxes are in the NIR wavelengths and lie above the detection limit of JWST's NIRCam and Roman's WFI.}
    \label{fig:flux}
\end{center}
\end{figure*}

\noindent
So far we have established that the emission from Pop \RomanNumeralCaps{3} TDEs mostly resides in UV/optical wavebands in the rest frame of the TDE host galaxy. In the subsequent calculations, we adopt a canonical value of redshift  $z=10$ for Pop \RomanNumeralCaps{3} star TDEs and calculate the red-shifted TDE emissions observed at $z=0$. 

With $\nu_{\rm e}$ and $\nu_{\rm o}$ denoting the rest-frame frequency and the observed frequency, respectively, the two are related as: 
\begin{equation}
    \nu_{\rm e} = (1+z)\nu_{\rm o}.
\end{equation}
Therefore, the observed Pop \RomanNumeralCaps{3} star TDE SED should be redshifted following $\nu_o = \nu_e/(1+z)$ and peaks around or slightly below $10^{14}$ Hz, which corresponds to $\lambda\sim10^3$ nm. This means that a large fraction of the Pop \RomanNumeralCaps{3} star TDE emission should be redshifted to the near-infrared (NIR) regime, and such events can be potentially detected by the Near Infrared Camera (NIRCam) on JWST covering the wavelength range of $600-5000$ nm as well as the Wide Field Instrument (WFI) on Roman with a waveband of $480-2300$ nm.

The observed specific flux, $F_\nu (\nu_{\rm o}$), is related to the rest-frame specific luminosity, $L_\nu (\nu_{\rm e})$, following \citep{Hogg02}
\begin{equation}
    F_\nu (\nu_{\rm o})= \frac{1+z}{4 \pi D_L^2}L_\nu (\nu_{\rm e}).
\label{eq:observedflux}
\end{equation}
Here $D_L$ denotes the luminosity distance which can be expressed as 
\begin{equation}
    %D_L = \frac{cz}{H_0} (1+z),
    D_L = (1+z)D_c,
\end{equation}
where $$ D_c=\frac{c}{H_0}\int_0^z\frac{dz}{\sqrt{\Omega_M(1+z)^3 + \Omega_\Lambda}}$$ is the comoving distance, $\Omega_M, \Omega_\Lambda, H_0$ is the matter, energy density parameter and Hubble constant at the current time. %=69.6~\rm km~s^{-1}Mpc^{-1}$ .

\begin{figure*}
\begin{center}
     \includegraphics[width=16cm]{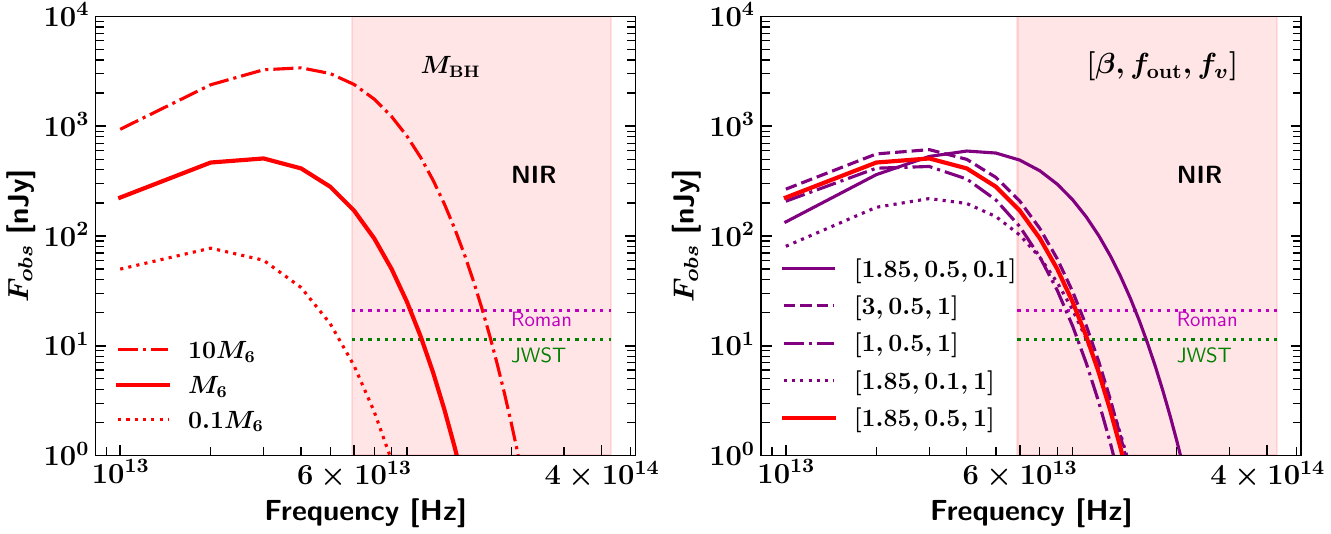}
    \caption{The dependence of the observed flux at $t=t_{\rm edge}$ on other parameters, such as the MBH mass ($M_{\rm BH}$), penetration factor ($\beta$), mass outflow fraction ($f_{\rm out}$) and wind velocity ($f_v$). Here we stick to the fiducial model of a Pop \RomanNumeralCaps{3} star with $M_\star=300M_\odot, Z=10^{-5}$. The shaded regions, magenta and green dotted lines are the same as in Fig. \ref{fig:flux}.}
    \label{fig:par}
\end{center}
\end{figure*}

We calculate the observed fluxes of Pop \RomanNumeralCaps{3} TDEs at different epochs throughout the event for different stellar models using Eq. \ref{eq:observedflux}. The results are plotted in Fig. \ref{fig:flux}. It can be seen that for all models the fluxes in the NIR band stay above the detection limit of the JWST NIRCam ($\sim 11$ nJy corresponding to the F150W filter) and Roman WFI ($\sim 20.9$ nJy corresponding to F106 filter) throughout these epochs. Also no significant differences in the flux level are observed due to different masses or metallicities of Pop \RomanNumeralCaps{3} stars. Moreover, we note that the observed flux from the TDE of a disrupted solar-type star is also detectable by NIRcam and WFI.

Additionally, we have examined the dependence of observed flux on other model parameters and the results are shown in Fig. \ref{fig:par}. Consistent with the results illustrated in Fig. \ref{fig:SED} (c) and (d), one can see that the NIR flux increases significantly with increasing $M_{\rm  BH}$ and moderately with decreasing $f_v$, but is barely affected by $f_{\rm out}$ and $\beta$. In particular, the NIR flux becomes undetectable even around peak luminosity when $M_{\rm BH}<10^5 M_\odot$, unless the wind is exceptionally slow ($f_v \lesssim 0.1$). In summary, in all models, the observed Pop III star TDE NIR flux around flare peak stays significantly above the detection limit of NIRCam and WFI, as long as $M_{\rm BH}\gtrsim 10^6 M_\odot$. Therefore, our results demonstrate that the detection of Pop \RomanNumeralCaps{3} star TDEs at $z\sim10$ with both JWST or Roman is feasible.

\begin{figure}
     \includegraphics[width=8.8cm]{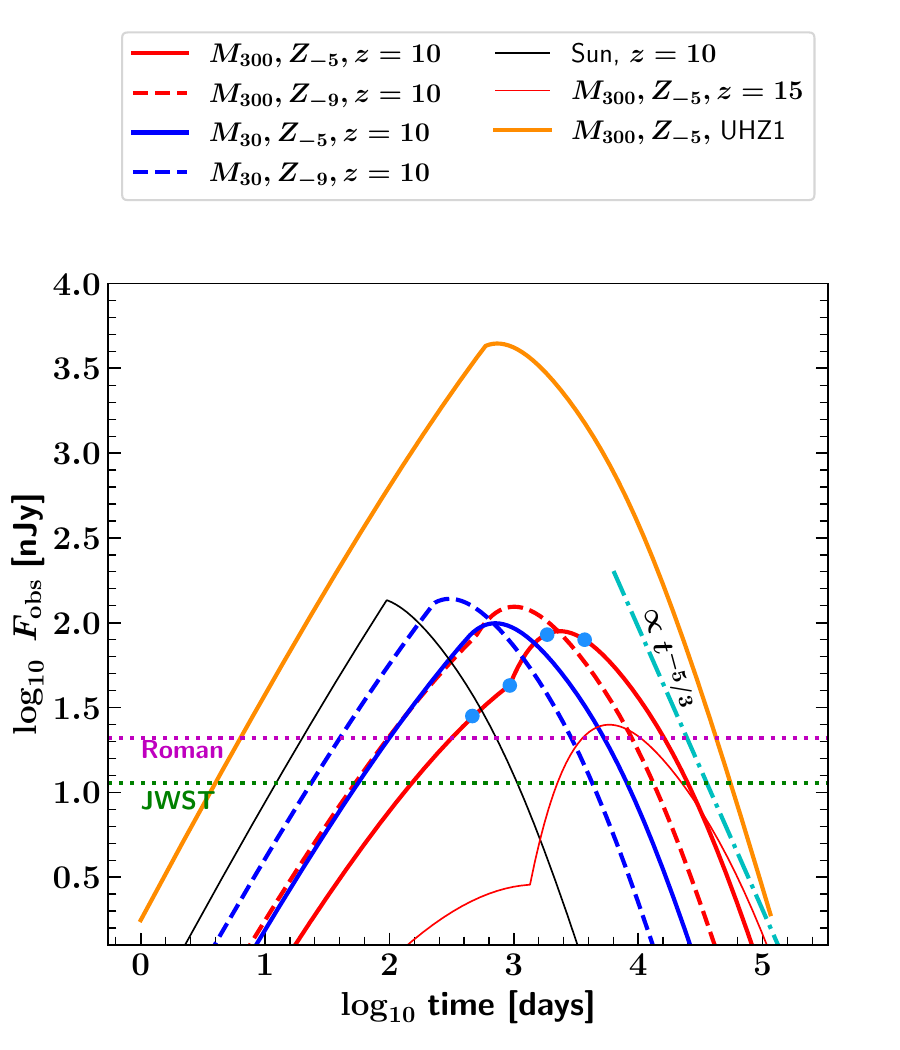}
    \caption{We show the NIR light curves of Pop \RomanNumeralCaps{3} star TDEs in comparison to that of a solar-type star. The NIR flux integrated are in the JWST NIRcam band. All TDEs happen around a MBH with $M_{\rm BH} = 10^6M_\odot$ at $z=10$. Color schemes and line styles are the same as Fig. \ref{fig:lc}. Furthermore, for the default model of Pop \RomanNumeralCaps{3} star ($M_\star=300M_\odot=M_{300},~Z=10^{-5}=Z_{-5}$), we also show the observed NIR light curve when the TDE happens at $z=15$ (red thin line) or when the MBH is a UHZ1-like source with $M_{\rm BH} = 4\times 10^7M_\odot$ at $z=10.1$ (orange line). The late-time light curves are compared to a power-law decay pattern of $t^{-5/3}$ (cyan dot-dashed line). Green and magenta dotted solid lines denote the flux limit of NIRCam (F150W) and WFI (F106) respectively.  It is noticed that observed NIR light curves evolve very slowly due to cosmological time dilation.}
    \label{fig:lc_obs}
\end{figure}

Next we calculate the NIR light curves of Pop \RomanNumeralCaps{3} star TDEs as would be observed in the JWST NIRcam band. The NIR light curves are shown in Fig. \ref{fig:lc_obs}, where one can see that the NIR flux can exceed both the JWST NIRCam and Roman WFI detection limits even when the TDE originates at $z=15$. Moreover, at such high redshifts, the time dilation effect due to cosmological redshift is very strong, since $t_{\rm o}=(1+z) t_{\rm e}$. Taking a Pop \RomanNumeralCaps{3} star TDE at $z=10$ as an example, the observed NIR light curve (above the detection limit) rises on timescales $\sim100-1000$ days and decays over $10^3-10^4$ days. At late time, the NIR flux still decays following a pattern relatively close to the classical mass fallback rate ($\propto t^{-5/3}$). 

Furthermore, inspired by the recent discovery of UHZ1, we include the case of a Pop \RomanNumeralCaps{3} star disrupted by a MBH with $M_{\rm BH} = 4\times 10^7M_\odot$ hosted by galaxy at $z=10.1$ (orange line). As expected, the observed NIR flux is significantly higher in this case, which increases the chance of observing such events in current and upcoming NIR surveys.

Interestingly, given the observed Pop \RomanNumeralCaps{3} star TDE flares rise over a few years, there is still a prospect of identifying such TDEs detected during the rising phase as transients, if multiple detections are made within the typical operation time of surveys. However, during the flare decay phase, as the evolution timescale is very long, these events will have almost near constant brightness during the limited operation time and therefore they are more likely to be mis-categorized as AGNs based on their photometry. One promising method to distinguishing Pop \RomanNumeralCaps{3} star TDEs from TDEs of Pop I stars or AGNs is through spectroscopic follow-ups.  Metal lines should be present in the spectra of either Pop I star TDEs or AGNs even at $z\sim10$ given that gas is likely no longer metal free \citep{Yang23}. However, if a Pop \RomanNumeralCaps{3} star TDE occurs in a previously quiescent galaxy, a metal-free spectrum should be produced, since all gas supplied to the MBH likely originates from the tidally disrupted Pop \RomanNumeralCaps{3} star. 

\subsection{Rates of Pop \RomanNumeralCaps{3} Star TDEs} \label{sec:rate}

\begin{figure}
    \centering
    \includegraphics[width=8.7cm]{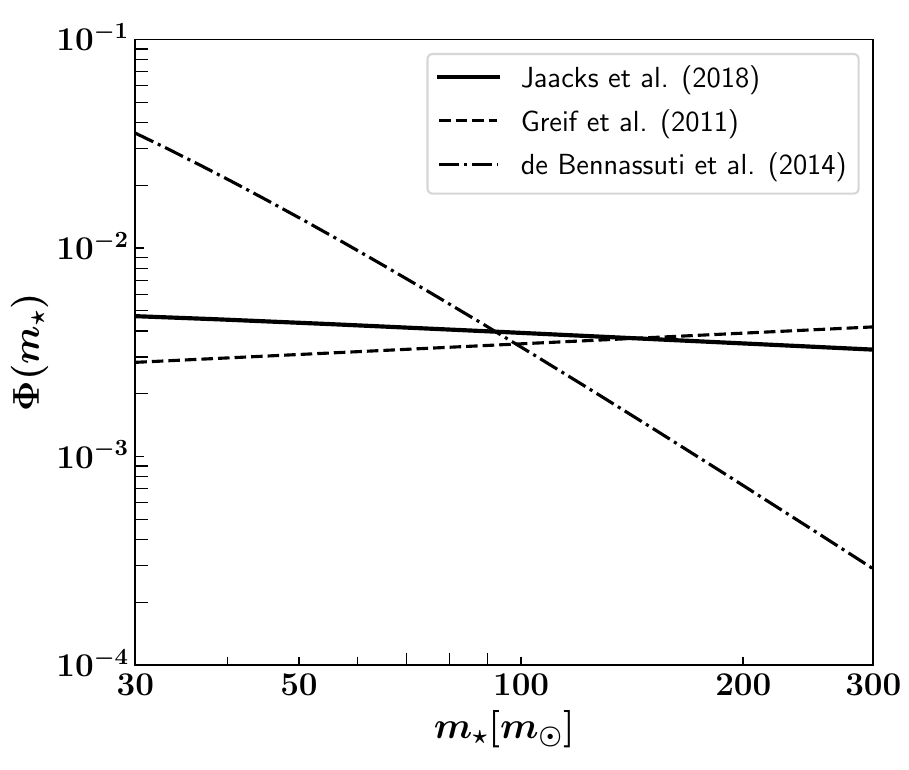}
    \caption{Different forms of IMF for the Pop \RomanNumeralCaps{3} stars as listed in Table \ref{tab:xx}.  It should be clarified that \cite{Jaacks18} calculated Pop \RomanNumeralCaps{3} IMF only within the mass range of $1M_\odot-150 M_\odot$, and  \cite{Greif11} calculated the IMF only within the mass range of $0.1M_\odot -10 M_\odot$. For this study, we extrapolated all three IMFs to the range of $30M_\odot -300 M_\odot$.}
    \label{fig:imf}
\end{figure}

\noindent
We now estimate the intrinsic rates of Pop \RomanNumeralCaps{3} TDEs in the early Universe. We adopt the approach in \cite{Pfister20, Pfister22} to calculate the TDE rate through two-body scattering. The differential TDE rate is given by: 

\begin{equation}
\begin{split}
    \frac{d^2\Gamma}{d\log m_\star d\log\beta} =  8 \pi^2 G M_{\rm BH}\frac{R_T}{\beta} \Phi(m_\star) m_\star  \\ \times \int_{0}^{E_m} \zeta (q,f,\beta,m_\star) \, dE\
    \label{eq:rate}
    \end{split}
\end{equation}
Here $\Phi (m_\star)$ is the stellar initial mass function (IMF),
$E$ is the specific orbital energy of the star with the maximum value at $E_m= GM_{\rm BH}/R_T$ corresponding to the orbit with $R_T$ as the closest approach, and $\zeta(q,f, \beta, m_\star)$ is a function of the loss-cone filling factor $q$,  the stellar distribution function $f$, the stellar mass $m_\star$ and the penetration parameter $\beta$. The TDE of a Pop \RomanNumeralCaps{3} star can be induced by the scattering between a Pop \RomanNumeralCaps{3} star and a normal star or that between two Pop \RomanNumeralCaps{3} stars. However, at the redshift range that we are considering, Pop \RomanNumeralCaps{3} stars contribute to only a few percent of total stellar mass \citep{Magg22}. 
Therefore, we only include the scattering of Pop \RomanNumeralCaps{3} stars by normal stars into our calculation. Further details of the terms in Eq. \ref{eq:rate} and the rate calculation are given in Appendix \ref{app:rateCalculation}.

\begin{table*} 
    \centering
	\begin{tabular}{cccccc} % four columns, alignment for each
		\hline
		IMF & $\Phi(m_\star)$ & $\alpha$ & $m_c$ & $\Gamma [\rm gal^{-1}yr^{-1}] $ & $\dot{N}_{\rm max} [\rm Mpc^{-3} yr^{-1}]$\\
		\hline
        \cite{Greif11} & $\propto m_\star^{-\alpha}$ & $-0.17$ & $-$ & $1.8\times 10^{-8}$ &$1.1\times 10^{-9}$\\ %$2.5\times 10^{-6}$\\ %$4.3\times 10^{-6}$\\ 
        %\cite{Stacy13} & $m_\star^{-\alpha}$ & $0.17$ & $-$ & $4.2\times 10^{-6}$\\
		\cite{deBennassuti14} & $\propto m_\star^{\alpha-1}\exp{(-m_c/m_\star)}$ & $-1.35$ & $20 M_{\odot}$& $4.1\times 10^{-8}$& $2.5\times 10^{-9}$\\ %1.6\times 10^{-6}$\\ %$1.0\times 10^{-5}$\\
        \cite{Jaacks18} &  $\propto  m_\star^{-\alpha}\exp{(-m_{c}^2/m_\star^2)}$ & $0.17$ & $4.47 M_{\odot}$& $2.0 \times 10^{-8}$ & $1.2\times 10^{-9}$\\%$2.4 \times 10^{-6}$\\%$4.2 \times 10^{-6}$ & 
		\hline
	\end{tabular}
 \caption{Different IMFs of Pop \RomanNumeralCaps{3} stars suggested by different works in the literature. $\Phi(m_\star)$ describes the  IMF equation, with $\alpha$ and $m_c$ denoting the slope and cut-off mass, respectively. $\Gamma$ represents the rate of Pop III star TDE from a single galaxy with $M_{\rm BH}=10^6 M_\odot$. $\dot{N}_{\rm max}$ represents the Pop III star TDE volumetric rate computed using the BHMF from the the merger-driven model. }
 \label{tab:xx}
\end{table*}

\begin{figure*}
    \centering
    \includegraphics[width=18.2cm]{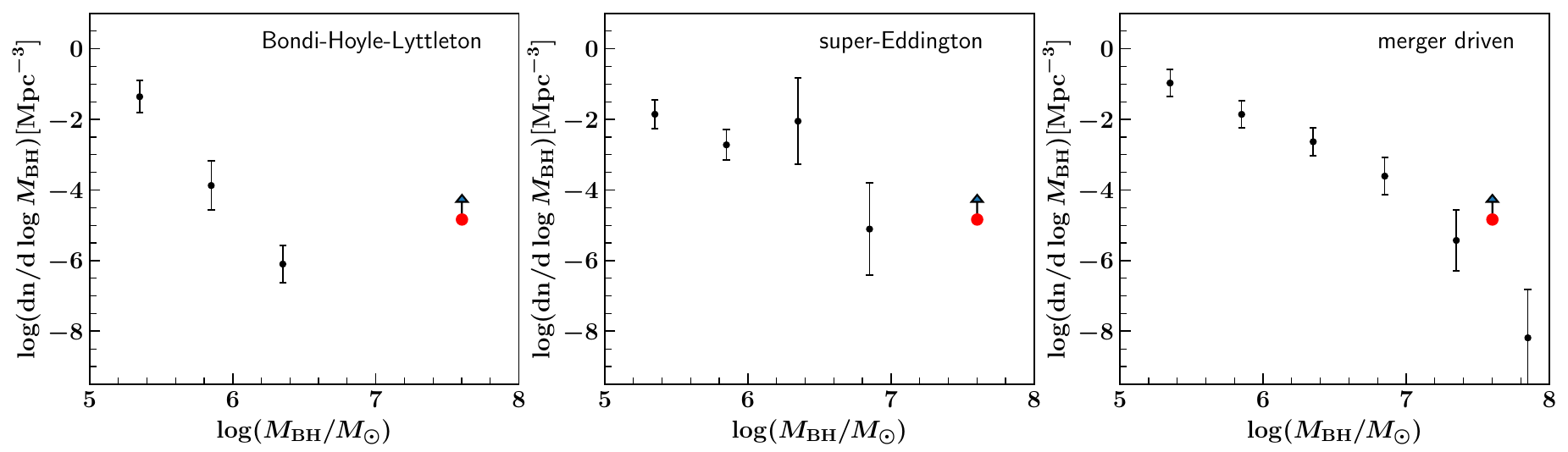}
    \caption{The black hole mass function at $z=10$ for different growth models (left: Bondi-Hoyle-Lyttleton; middle: super-Eddington; right: merger driven). The BHMFs are adopted from \cite{Trinca22} with permission. The red points in all the three panels denote the BHMF inferred from the detection of UHZ1.}
    \label{fig:bhmf}
\end{figure*}

We summarize the assumptions used for the TDE rate calculation as below:

\begin{itemize}
    \item Ranges of parameters: We adopt the MBH mass range of $M_{\rm BH} =[10^5M_\odot,10^8 M_\odot]$, Pop \RomanNumeralCaps{3} stellar mass range of $m_\star=[30M_\odot,300M_\odot$], normal star in the range of  $m_\star=[0.1M_\odot,10M_\odot$], and $\beta$ in the range $[\beta_c=1.85, R_T/R_s]$ to calculate the total TDE rate by integrating Eq. \ref{eq:rate}.
    
    \item IMF of the stellar populations: We adopt  the general Kroupa IMF \citep{Kroupa01} for the normal main-sequence stars.  
    However, the IMF for Pop \RomanNumeralCaps{3} stars is model-dependent and not constrained very well yet. Different IMFs have been proposed in the literature, and a few examples are shown in Table \ref{tab:xx} and Fig. \ref{fig:imf}. However, using different Pop \RomanNumeralCaps{3} star IMFs bring negligible effects to TDE rates, as explained in Appendix \ref{app:rateCalculation}. 
    
    \item Mass fraction of Pop \RomanNumeralCaps{3} stars: \cite{Magg22} showed that at $z\sim10$ Pop \RomanNumeralCaps{3} stars has a mass fraction between $1\%$ and $15\%$ among all stellar population depending on the timescale at which  Pop \RomanNumeralCaps{3} stars transit to Pop \RomanNumeralCaps{2} stars. We adopt a Pop \RomanNumeralCaps{3} star mass fraction of $7\%$ throughout the galaxy for our calculation.
    
    \item Stellar density distribution: As there lacks studies of the structures of galaxies for different stellar populations at these extremely high redshifts, we follow the classical papers such as \cite{WangMerritt04, Stone16, Pfister20} and assume isothermal stellar density distributions in the Keplerian potential for both Pop \RomanNumeralCaps{3} and normal stars, which gives 
    \begin{equation}
    \label{eq:iso}
    \rho(r) = \rho_{0}(r/r_{\rm inf})^{-\alpha} 
    \end{equation}
    where $\rho(r)$ is the stellar density, $\alpha =2$, and $r_{\rm inf} = GM_{\rm BH}/2\sigma^2$, with $\sigma$ being the velocity dispersion of the host galaxy.

\end{itemize}

Under these assumptions, the TDE rate of Pop \RomanNumeralCaps{3} stars from a single galaxy hosting $10^6M_{\odot}$ MBH is calculated and listed in Table \ref{tab:xx}. This intrinsic TDE rate is $\Gamma\sim10^{-8}~{\rm gal^{-1}~yr^{-1}}$, which is barely affected by the choice of Pop \RomanNumeralCaps{3} star IMF. 

\begin{figure}
    \centering
    \includegraphics[width=8.5cm]{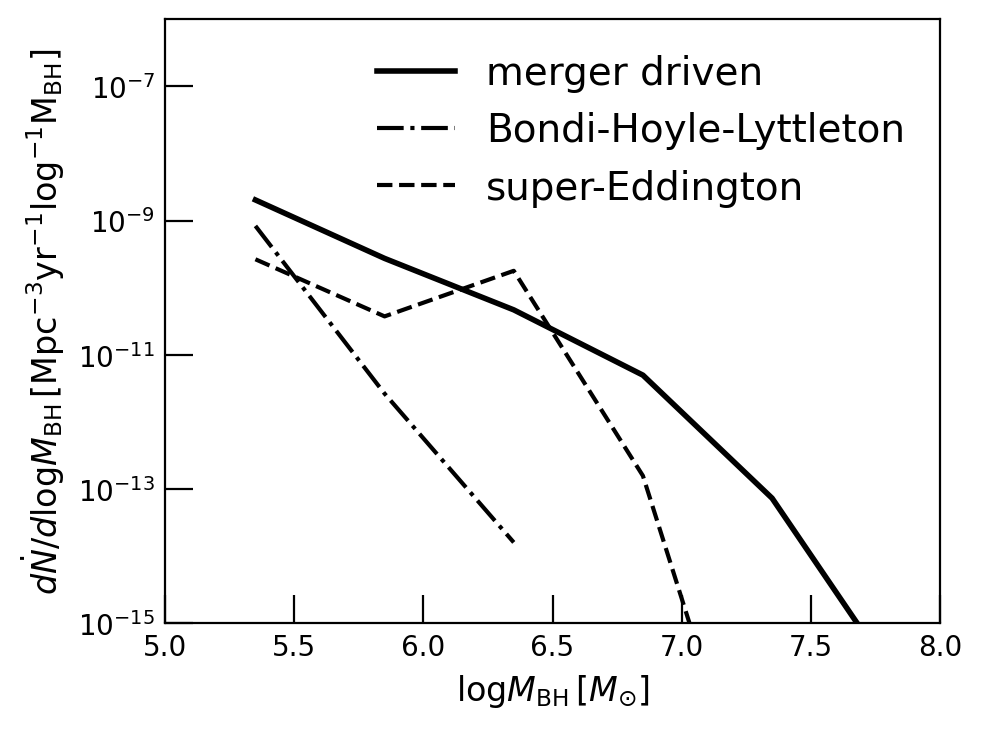}
    \caption{The differential Pop III star TDE volumetric rates as a function of $M_{\rm BH}$ for the three different BHMFs in Fig. \ref{fig:bhmf}. One can see that the BHMF through the merger-driven growth channel overall produces the highest TDE volumetric rate.}
    \label{fig:volumetric_rate}
\end{figure}

One can further compute the volumetric rate of Pop \RomanNumeralCaps{3} star TDEs using: 
\begin{equation}
    \begin{split}
        \dot{N} = \iiint \frac{dn}{d\log M_{\rm BH}}  \times  \frac{d^2\Gamma}{d \log m_\star d \log \beta} \\
        \times d \log m_\star~ d\log \beta~ \, d\log M_{\rm BH}
    \end{split}
\end{equation}
here $\Phi({M_{\rm BH}})=dn/d\log M_{\rm BH}$ denotes the black hole mass function (BHMF), which is defined as the number of MBHs with masses between $\log M_{\rm BH}$ and $\log M_{\rm BH}+d\log M_{\rm BH}$ in unit co-moving volume. We note that the BHMF at high redshifts sensitively depends on the seeding scenarios and growth channels of MBHs, much of which is currently uncertain. In this work, we adopt three different BHMFs from \cite{Trinca22}, that consider not only light, intermediate and heavy MBH seeds but also different processes responsible for their growth. Fig. \ref{fig:bhmf} shows the BHMF at $z=10$ for these three growth models, namely, Eddington-limited Bondi-Hoyle-Lyttleton accretion, super-Eddington accretion and merger-driven growth \citep{Hoyle41, Bondi52}. We focus on the MBHs with $M_{\rm BH} \geq 10^5M_\odot$ for this work, since the observed Pop \RomanNumeralCaps{3} star TDE flux goes below the threshold of NIRCam and WFI when $M_{\rm BH} < 10^5M_\odot$ (Fig. \ref{fig:par}). It is seen that the BHMF from the merger-driven growth channel generally has higher values in this $M_{\rm BH}$ range than the other two BHMF choices. The volumetric Pop \RomanNumeralCaps{3} star TDE rates using these three different BHMFs are calculated and shown in Fig. \ref{fig:volumetric_rate}, which are around $10^{-10}-10^{-9} \rm Mpc^{-3}yr^{-1}$. Out of the three, not surprisingly, the BHMF based on the merger-driven growth model produces the highest integrated volumetric TDE rate $\dot{N}_{\rm max} \approx 10^{-9} \rm Mpc^{-3}yr^{-1}$.

 We also include the BHMF inferred from calibrating models with the detection of UHZ1 in Fig. \ref{fig:bhmf}, and note that the BHMF taken from \cite{Trinca22} (as well as other current BHMF models) underestimate the number density of massive black holes in all the growth models. Hence it is possible that the actual Pop III TDE rate is higher than our estimate. An updated BHMF at high redshift from additional observations of a population of $z \sim 10$ BHs will be crucial for more accurately constraining the Pop III star TDE rate.

\subsection{Pop \RomanNumeralCaps{3} Star TDE Predictions for JWST and Nancy Grace Roman Space Telescope} \label{sec:number}

\noindent
In this section we estimate the total number of Pop \RomanNumeralCaps{3} star TDEs that stand to be detected by telescope JWST and Roman. For a first-order estimate of the upper limit of the detection numbers, we ignore factors such as survey strategies and limitations. The total number can then be approximated using:
\begin{equation}
    N = \dot{N} \times V \times T.
    \label{eq:obs_rate}
\end{equation}
Here $\dot{N}$ is the volumetric TDE rate obtained in the previous section, $T$ is duration of the observational survey, and $V$ is the co-moving volume from which the event can be detected. $V$ can be computed using the following equation:  
\begin{equation}
V= \frac{4\pi [d(z_{\rm max})^3-d(z_{\rm min})^3]}{3} \times f_{\rm sky},
\end{equation}
where $d(z)$ is the co-moving distance dependent on redshift $z$, and $f_{\rm sky} $ is the sky coverage fraction of the telescope of survey.

For the case of Pop \RomanNumeralCaps{3} star TDEs, we adopt $z_{\rm min}=10$, when the Pop \RomanNumeralCaps{3} star population reaches peak, and $z_{\rm max}=15$, when both the Pop \RomanNumeralCaps{3} stars and MBHs should have started to form. More importantly, we have predicted in Fig. \ref{fig:lc_obs} that Pop \RomanNumeralCaps{3} star TDEs at $z=15$ are still bright enough to be detected by JWST and Roman. Moreover, for simplicity, we stick to a constant $\dot{N}=1.2 \times 10^{-9} \rm \ Mpc^{-3}\ yr^{-1}$ (based on the Pop \RomanNumeralCaps{3} star IMF by \cite{Jaacks18} and the BHMF from the merger-driven growth model), and note that this likely gives an upper limit to the detected number. 

JWST Cycle 1 has various deep surveys, such as COSMOS-Web with a total area of 0.54 $\rm deg^2$ and duration of 255 hours \citep{Casey23}, and  the JWST Advance Deep Extragalactic Survey (JADES) with its sky coverage of $ 236 ~\rm arcmin^2$ and a total duration of $426$ hours considering both the deep and medium modes \citep{Eisenstein23}. Using both the COSMOS-Web survey and the deep mode of the JADES survey, which can both probe galaxies at $z> 10$, the total detected number of Pop \RomanNumeralCaps{3} TDEs in one year is about $5\times 10^{-4}$. Throughout the 10 year expected lifespan of JWST, even if we assume COSMOS-Web like surveys are continuously conducted, the total expected detection number is still only $(10 \ {\rm yr}/ 255 \ {\rm hr}) \times 5\times 10^{-4} \sim 0.2 $ event in total. Therefore, the chance of detecting Pop \RomanNumeralCaps{3} star TDEs using JWST is slim. If such TDEs were detected, it would indicate the BHMF at $z>10$ is much larger than currently estimated, which would pose an interesting challenge in our understanding of the formation efficiency and evolution of MBHs in the early Universe. With the recent discovery of UHZ1 and its interpretation as an OBG arising from a heavy initial seed by \cite{Natarajan+2023}, it appears that BH seed formation in the early Universe can occur via multiple seeding pathways hence rendering the process significantly more efficient than previously believed. 

A much more promising telescope for detecting Pop \RomanNumeralCaps{3} star TDEs is the upcoming Nancy Grace Roman Space Telescope, which is designed to have a sensitivity similar to JWST (Fig. \ref{fig:lc_obs}) and conduct wide-field surveys over a very large FOV of $0.281 ~\rm deg^2$ \citep{Mosby20}. Hence, from the High Latitude Wide Area Roman survey that aims to cover a total sky area of $\sim 2000$ deg$^2$ \citep{WangY22}, the number of detected Pop \RomanNumeralCaps{3} TDEs can reach $\lesssim 60$ in one year. Furthermore, there is another proposed survey, the Next-generation All-sky Near-infrared Community surveY (NANCY) \citep{Han23}, which plans to perform an all-sky scan. This survery strategy will lead to the detection of $\sim 1000$ Pop \RomanNumeralCaps{3} TDEs in a single year. Once a large sample of Pop III star TDEs are observed, the number can be used to put a strong constraints on the properties and mass fraction of Pop III stars as well as the BHMF at $z\gtrsim10$ which will provide the much needed insights into the efficiency of early BH formation.

\section{Summary and Discussion}\label{sec:discussion}

\begin{figure*}
\begin{center}
     \includegraphics[width=18.cm]{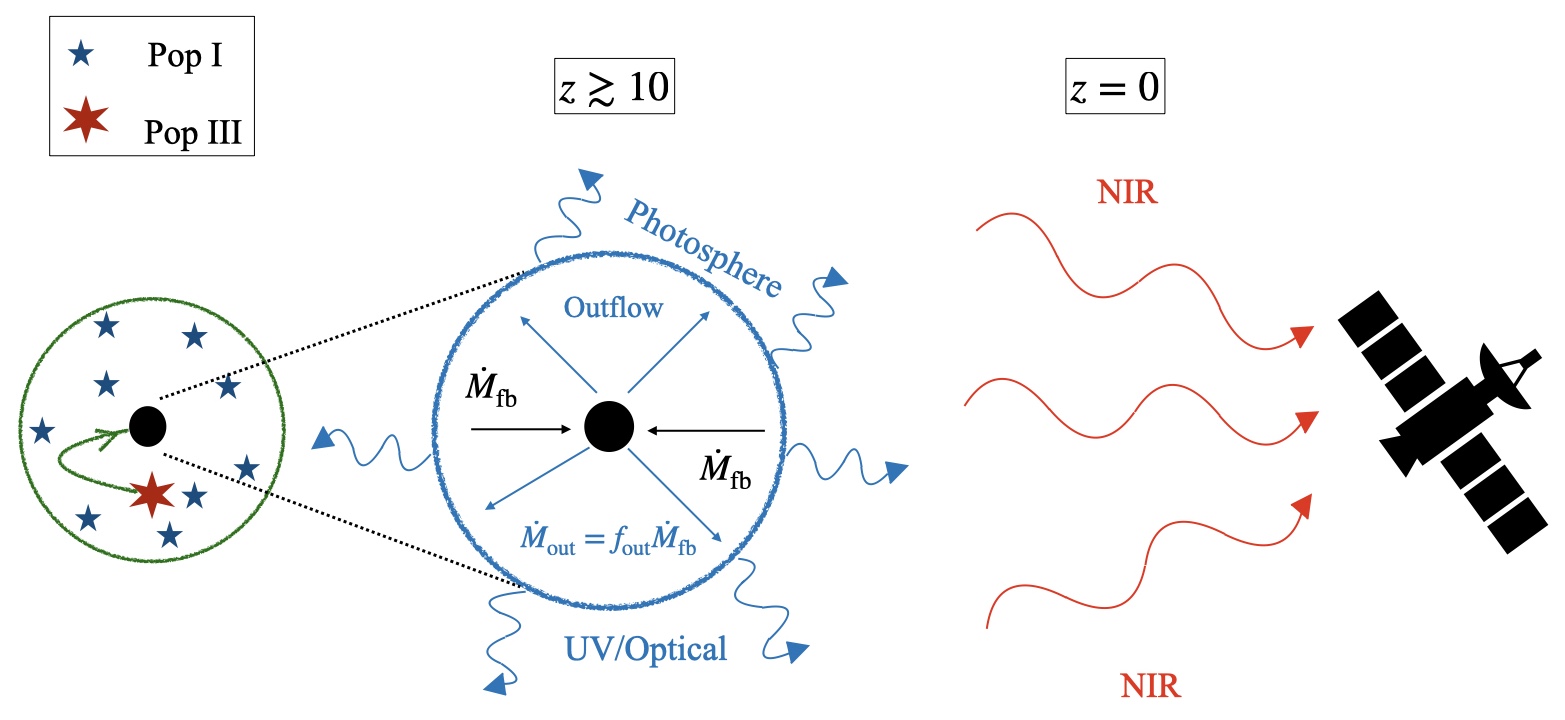}
    \caption{Schematic diagram of Pop \RomanNumeralCaps{3} TDEs. In a galaxy at $z\gtrsim10$, a Pop \RomanNumeralCaps{3} star is scattered by other stars and gets tidally disrupted by the MBH. A large fraction of the fallback debris material is converted to a powerful wind with outflowing mass rate $\dot{M}_{\rm out}=f_{\rm out}\dot{M}_{\rm fb}$, which produces a luminous optical/UV flare. At $z=0$, the flare emissions are redshifted to the NIR wavelength and can be detected by JWST and the Nancy Roman Space Telescope.}
    \label{fig:schematic}
\end{center}
\end{figure*}

\noindent Pop \RomanNumeralCaps{3} stars are the first-generation stars which are expected to form from metal-free primordial gas. Their formation and properties have been under study with multiple theoretical investigations, but their direct detection remains a challenging open question. The primary focus of this work is to investigate the prospect of detecting Pop \RomanNumeralCaps{3} stars through the flares produced when these stars are tidally disrupted by MBHs, utilizing telescopes such as JWST and Roman that  can probe the high redshift Universe. We summarize the key features of our assumed model and main findings therefrom below and show a schematic illustration in Fig. \ref{fig:schematic}.

\begin{itemize}
    \item Pop \RomanNumeralCaps{3} stars have tidal disruption radii around or a few times larger than that of a solar-type star. Very massive Pop \RomanNumeralCaps{3} stars with $M_\star>100M_\odot$ can possibly be disrupted by MBHs with $M_{\rm BH}$ up to $\sim10^9 M_\odot$.

    \item The debris mass fallback rates for Pop \RomanNumeralCaps{3} star TDEs can stay hyper-Eddington for long periods. For example, for a typical MBH with $M_{\rm BH}=10^6 M_\odot$, $\dot{M}_{\rm fb}$ can reach a peak value of $10^4-10^6 \dot{M}_{\rm Edd}$ and stays super-Eddington for $10-100$ years. In the more extreme case of a UHZ1-like source with $M_{\rm BH} \gtrsim 10^7 M_\odot$, $\dot{M}_{\rm fb, peak}$ drops to around $100-10^5 \dot{M}_{\rm Edd}$ and $\dot{M}_{\rm fb}$ can still stay super-Eddington for more than a few years.
    
    \item We adopt the super-Eddington outflow model proposed by \cite{Strubbe09} and only consider the emission produced by the outflow in this phase while predicting potential observational signatures.
    
    \item In the rest frame of the host galaxy, Pop \RomanNumeralCaps{3} star TDEs mainly produce UV/optical emission (Fig. \ref{fig:SED}), with the bolometric luminosity increasing almost linearly with $M_{\rm BH}$. These tidal flares have much longer evolution timescales compared to typical Pop I star TDEs detected currently (Fig. \ref{fig:lc}). 
    
    \item  As Pop \RomanNumeralCaps{3} star TDEs are expected to mostly occur at high redshifts ($z\gtrsim 10$), a large fraction of the emission is redshifted to the NIR band and the observed flux exceeds the detection limit of JWST NIRCam and Roman WFI (Fig. \ref{fig:flux} and Fig. \ref{fig:par}). Moreover, due to the time dilation effect, the observed NIR flares evolve even more slowly (rising in 100-1000 days and decaying over 10-100 years) (Fig. \ref{fig:lc_obs}).
    \item The volumetric rate of Pop \RomanNumeralCaps{3} star TDEs is insensitive to Pop \RomanNumeralCaps{3} star IMF, but sensitively depends on the BHMF at high redshifts which in turn depends on the seeding and growth models of MBHs. 
    \item We obtain an upper limit on Pop \RomanNumeralCaps{3} star TDE volumetric rate of $\sim 10^{-9} \rm Mpc^{-3}~yr^{-1}$ (Fig. \ref{fig:volumetric_rate}), based on the current BHMF model. We note that the recent detection of overmassive black holes at high redshifts indicate that the density of MBHs at $z\sim10$ and hence our estimated TDE rates could be lower than the actual numbers. 
    \item The high NIR luminosity and long duration of the Pop \RomanNumeralCaps{3} star tidal flares make these events detectable by the Roman Space telescope and we predict up to a few tens of events per year with its wide field instrument. However, the probability of detecting a Pop \RomanNumeralCaps{3} TDE using JWST is low due to its narrower FOV.
\end{itemize}

Our results under the SQ09 model are rather robust against the choice of Pop \RomanNumeralCaps{3} star mass, metallicity, and other model parameters such as the penetration parameter of the stellar orbit, outflow fraction and speed, etc. However, it should be acknowledged that several analytical models for TDE super-Eddington outflows have been proposed in literature, which is mentioned in Appendix \ref{app:comparison2}. Also very importantly, we note that this 1D analytical model by SQ09 cannot explain the observed X-ray emissions produced in Pop I star TDEs in the local universe. For the latter, simulations of super-Eddington disks around SMBHs with $\dot{M}_{\rm acc}\sim (1-10)\dot{M}_{\rm Edd}$ reveal that an optically thin funnel can form in the polar region through which X-rays can leak out \citep{Dai18, Jiang19, Thomsen22}. However, it is previously shown that the funnel can diminish when $\dot{M}_{\rm acc}$ reaches $>100\dot{M}_{\rm Edd}$ \citep{Sadowski16}. Therefore, Pop \RomanNumeralCaps{3} star TDEs with extremely high fallback rates likely cannot produce strong X-ray emissions, unless when relativistic jets are produced and pointing towards the observer. The production of jets under the context of super-Eddington accretion is still being actively explored \citep[e.g.][]{CB20, Ricarte23}, and the rates of jetted Pop III star TDEs can be used to probe the MBH spin distributiuon in the early universe if the jet is magnetically driven \citep{BZ77}. Moreover, we note that the realistic structures of the massive stars can affect the TDE debris mass fallback rate calculation \citep{LS20}. Therefore, future detailed modeling of Pop \RomanNumeralCaps{3} star structures will be useful for improving our calculation.

As Pop \RomanNumeralCaps{3} stars are short-lived, a fraction of them should have undergone substantial nuclear burning at the time of disruption and therefore have an evolved, more extended structure, which will increase the chance of the star being tidally stripped or partially disrupted. In this work, we have already calculated the impact of the penetration parameter on the TDE fluxes Fig. \ref{fig:SED}(d) and \ref{fig:par}. As expected, the debris mass peak fallback rate will be reduced in the case of partial TDEs, but the flare flux can still be detected by the JWST and Roman for moderate $\beta$ values. Interestingly, recent studies have found that the rates of partial TDEs are significantly higher than the full disruption of solar-type stars in the local universe \citep{Zhong22, Bortolas23}. The same could apply for Pop III star TDEs, which should greatly enhance their detection rates.

It is also worth mentioning that at the high redshifts we are considering, Pop \RomanNumeralCaps{3} stars only occupy a few percent of the total stellar population, while Pop I and Pop II stars dominate the stellar mass and therefore should produce much higher TDE rates. Although the focus of this work is on Pop \RomanNumeralCaps{3} star TDEs, we have shown that the NIR flux of Pop I star TDEs at $z\sim10$ also exceeds the detection limit of JWST and Roman. Pop III star and Pop I star TDEs, however, can be distinguished by their different evolution timescales. Furthermore, the Pop I star TDEs observed so far typically produce metal lines including C, N, O, Mg, and Fe lines  in their optical or UV spectra \citep{Leloudas19, Blagor19, Gezari21, Chara22}, and such signatures cannot be produced by Pop \RomanNumeralCaps{3} star TDEs. Pop I and Pop II star TDEs at high redshifts are worth further investigation, which is beyond the scope of this paper.

Another interesting aspect worth considering is the detection of TDEs at high redshifts magnified through gravitational lensing. Recent works show promising results of finding extremely faint high-$z$ objects, including a number of Pop \RomanNumeralCaps{3} star candidates, faint stars, distant galaxies and black holes with their observed fluxes largely boosted by nearby foreground lensing clusters \citep{Kelly18, Kaurov19, Schauer22, vikaeus22a, Bogdan+2023, Meena23, Diego23, Chen24, Szekerczes24}. Hence, the effect of lensing should also enhance the observed TDE fluxes, which will bring more high-$z$ (Pop \RomanNumeralCaps{3}) TDEs into view. While it has been shown that the chance of having a high-$z$ galaxy lensed is roughly $\sim 10^{-3}$ \citep{Saha24}, the probability of having lensed transients yet to be studied in detail.

With the advent of JWST, the number of detected quasars with $M_{\rm BH}\gtrsim 10^9 M_\odot$ at $z>6$ has increased by a large number \citep{Wang23, Yang23, Natarajan+2023}, which offers interesting insights into the models of BH seed formation. TDEs including Pop \RomanNumeralCaps{3} TDEs could also be one of the channels responsible for the rapid growth of BH seeds at very high redshifts. This was previously proposed by \cite{Pfister21} who showed that intermediate mass black holes at high redshifts can accumulate their masses through TDEs at a similar rate as through gas accretion.

In summary, in this work we compute the properties of a brand new class of high-redshift sources, tidally disrupted Pop \RomanNumeralCaps{3} stars, and demonstrate that they could be viably detected in upcoming wide field IR surveys. The prospect of observing such TDEs has only become promising due to the recent launch of JWST and the expected launch of the Nancy Grace Roman Space Telescope.

\begin{acknowledgments}
We thank A. Amruth, A. Bera, C. Bottrell, T. Kwan, S. Kimura, T. Matsumoto, L. Thomsen, N. Yoshida, Z. Zhang for helpful discussions and A. Trinca for providing the data on BHMF. We also thank the referee for providing useful comments.
RKC, NYC and LD acknowledge the support from the National Natural Science Foundation of China and the Hong Kong Research Grants Council (12122309, N\_HKU782/23, 17304821, 17314822, 27305119).
PN acknowledges support from the Gordon and Betty Moore Foundation and the John Templeton Foundation that fund the Black Hole Initiative (BHI) at Harvard University where she serves as one of the PIs.
\end{acknowledgments}


\appendix

\counterwithin{figure}{section}
\section{Comparison between Models for the Fallback Time and Peak Fallback Rate} \label{app:comparison}

\noindent In this work, we have followed the results obtained by \cite{Guillochon13} to calculate the fallback time $t_{\rm fb}$  and peak fallback rate $\dot{M}_{\rm fb, peak}$ (Eq. \ref{eq:tfb} and Eq. \ref{eq:Mfbpeak} respectively) in TDEs. GR13 used the hydrodynamical simulation FLASH to calculate these parameters for different stellar structures assuming polytropic density distribution with $\gamma=4/3$ or $\gamma=5/3$. There are a few recent works which have also investigated the disruption process and the dependence of  $t_{\rm fb}$  and $\dot{M}_{\rm fb, peak}$ on stellar properties, such as mass and age, using simulations or semi-analytical calculations involving more accurate stellar structures obtained using MESA \citep[e.g.,][]{LS20, CN22, B24}. Here we include a direct comparison on $t_{\rm fb}$ and $\dot{M}_{\rm fb, peak}$ between the GR13 model and the results from \citet{B24} (B24 hereafter). 

\begin{figure}[ht]
\begin{center}
     \includegraphics[width=18.cm]{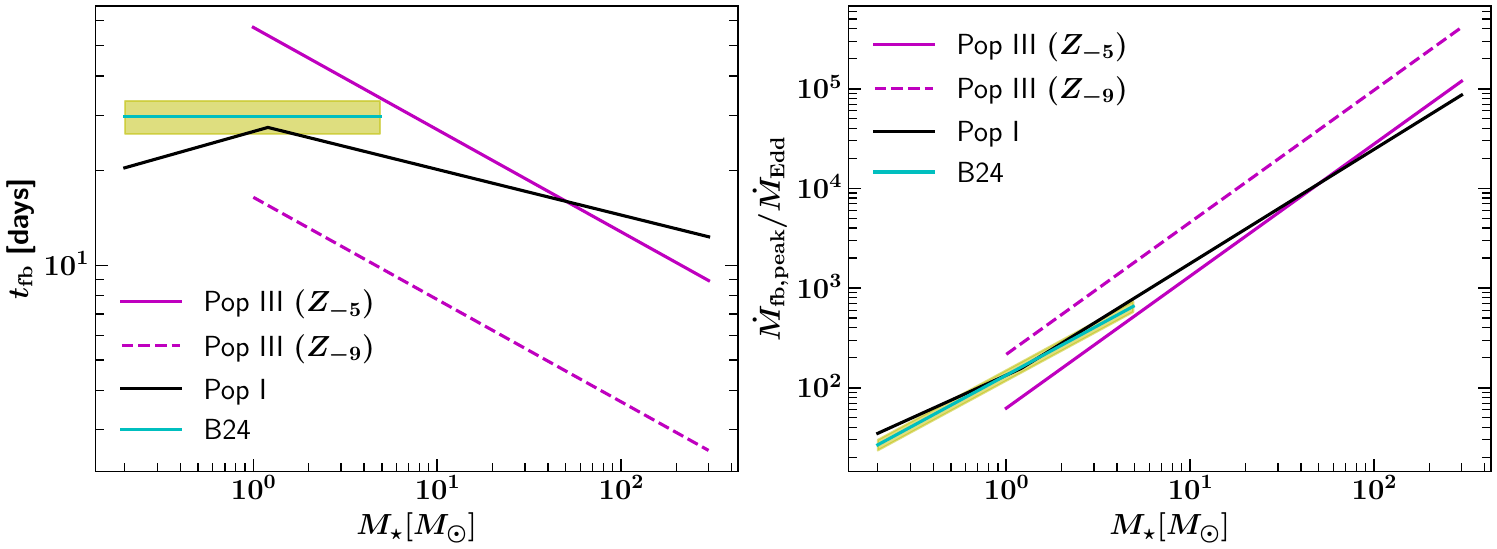}
    \caption{The fallback time $t_{\rm fb}$ and peak debris mass fallback rate $\dot{M}_{\rm fb, peak}$ as functions of stellar mass using both the GR13 model and the B24 model. The magenta solid and dashed lines represent TDEs of Pop \RomanNumeralCaps{3} stars with $Z=10^{-5}$ and $Z=10^{-9}$ respectively. The solid black line represents TDE of Pop \RomanNumeralCaps{1} stars. All of the lines mentioned above adopt the GR13 model. The cyan solid line represents the results from B24 for Pop \RomanNumeralCaps{1} stars (in the stellar mass range of $0.2-5M_\odot$). The yellow shaded regions denote the estimated uncertainties of the B24 model.}
    \label{fig:peak_ms}
\end{center}
\end{figure}

We first plot the $t_{\rm fb}$ and $\dot{M}_{\rm fb, peak}$ as a function of the stellar mass $M_\star$ in Fig. \ref{fig:peak_ms} for both Pop I and Pop III star TDEs. One can see that B24 indicates that $t_{\rm fb}$ is independent of the stellar mass and structure, although their results is obtained by studying TDEs of solar-metallicity stars in the mass range of $0.2-5M_\odot$ only (and therefore we plotted the B24 result in this mass range). Also, for Pop I stars,  $\dot{M}_{\rm fb, peak}$ obtained by GR13 and B24 are quite similar (in the mass range of $0.2-5M_\odot$).

\begin{figure}[ht]
\begin{center}
     \includegraphics[width=18.cm]{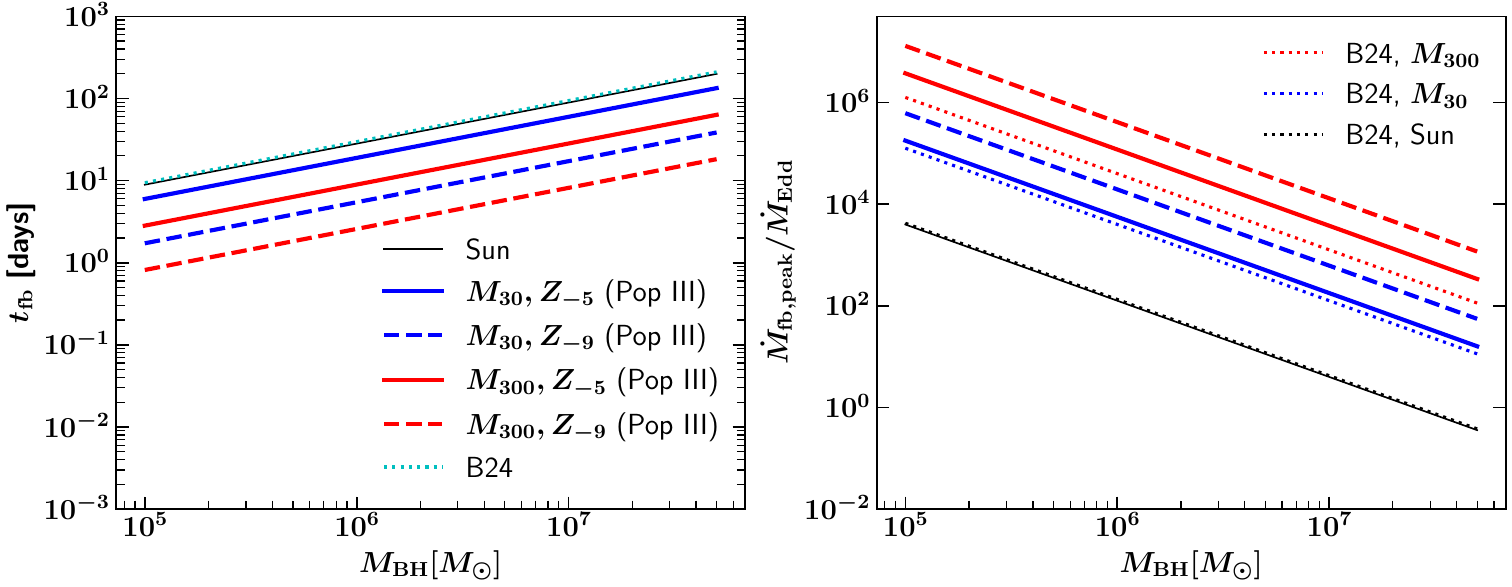}
    \caption{The  fallback time $t_{\rm fb}$ and peak debris mass fallback rate $\dot{M}_{\rm fb, peak}$ as functions of the MBH mass using both the GR13 model and the B24 model. The solid and dashed lines are same as Fig. \ref{fig:tpeak} (based on the GR13 model). The cyan dotted line in the left panel represents the fallback time using the B24 model. The black, blue and red dotted lines in the right panel correspond to $\dot{M}_{\rm fb, peak}$ in TDEs of a solar-type Pop I star, a $30 M_\odot$ Pop III star and a $300 M_\odot$ Pop III star respectively using the B24 model.}
    \label{fig:peak_ref}
\end{center}
\end{figure}

We further plot $t_{\rm fb}$ and $\dot{M}_{\rm fb, peak}$ as a function of $M_{\rm BH}$ in \ref{fig:peak_ref}, which is the same as Fig. \ref{fig:tpeak} except that we have added additional curves based on the results of B24. Here if we assume the peak fallback rate formula in B24 would also apply for Pop III star TDEs, then the peak mass fallback rates do not deviate much from those calculated using the GR13 formula for stars with different masses and metallicities.

\section{Calculation of TDE Rates} \label{app:rateCalculation}
\noindent For a distribution of Pop \RomanNumeralCaps{3} stars with density profile following $\rho \propto r^{-\alpha}$ and mass distribution expressed by $\Phi_{\rm PopIII}(m_\star)$, the TDE rate can be calculated using the loss cone dynamics. We give a summary of the methodology as below and refer the readers to  \citet{Pfister20, Pfister22} for more details. The differential TDE rate can be expressed in terms of stellar mass ($m_\star$), black hole mass ($M_{\rm BH}$) and the penetration parameter ($\beta$) such that:
\begin{equation}
     \frac{d^2\Gamma}{d \log m_\star d\log \beta} =  8 \pi^2 G M_{\rm BH}\frac{R_T}{\beta^2}  \Phi_{\rm Pop III}(m_\star) m_\star  \int_{0}^{E_m} \zeta(q(E,m_\star),f(E),\beta, m_\star)\, dE\ 
\end{equation}
The specific binding energy $E$
%$E=\phi(r) - v^2/2$, 
is integrated from 0 to the maximum energy $E_m=GM_{\rm BH}/R_T$. The function $\zeta(q,f,\beta,m_\star)$ is described as follows: 
\begin{align}
    \begin{split}
        \zeta(q,f,\beta,m_\star) ={} & \frac{f \log^2 e}{1+q^{-1}\xi(q) \ln (1/R_{\rm{lc}}(m_\star))} \bigg[1-2 \sum^{\infty}_{m=1} \frac{e^{-\alpha^2_m q/4}}{\alpha_m} \frac{J_0(\alpha_m \beta^{-1/2})}{J_1(\alpha_m)}\bigg]
    \label{eqt:G}
    \end{split}
\end{align}
where $R_{lc}$ describes the fraction of stars in the loss cone:
\begin{equation}
    R_{lc}(m_\star) = \frac{L_{\rm lc}^2}{L_{\rm circ}^2} \sim \frac{4E R_T(m_\star)}{GM_{\rm BH}}
\end{equation}
and 
\begin{equation}
    \xi(q) = 1-4\sum^\infty_{m=1} \frac{e^{-\alpha^2_m q/4}}{\alpha_m^2}
\end{equation}
$J_0$ and $J_1$ are the Bessel functions of the first kind, with general notation as:
\begin{equation}
    J_\alpha(x) = \sum^\infty_{m=0} \frac{(-1)^m}{m! \, \hat{\Gamma}(m+\alpha+1)}\bigg(\frac{x}{2}\bigg)^{2m+a}   
    \label{eqt:bessel}
\end{equation}
where $\hat{\Gamma}(x)$ is the gamma function, and $\alpha_m$ is the $m$th zero of the Bessel function $J_0$.

If we assume the density follows an isothermal profile with $\alpha=2$ (Eq \ref{eq:iso}), the distribution function $f(E)$ takes an analytical form:
\begin{equation}
    f(E) = (2\pi \sigma_{\rm inf}^2)^{-3/2} \frac{\rho_0}{\langle m_\star \rangle} \frac{\gamma (\alpha+1)}{\gamma(\alpha-1/2)}\bigg(\frac{E}{\sigma_{\rm inf}^2}\bigg)^{\alpha-3/2}
\end{equation}
where $\sigma_{\rm inf} = (GM_{\rm BH}/r_{\rm inf})^{1/2}$ is the velocity distribution at the BH influence radius $r_{\inf}$, $\langle m_\star \rangle$ is the average stellar mass of the Pop \RomanNumeralCaps{3} stellar mass function $\Phi_{\rm PopIII}(m_\star)$, $\gamma$ is the Euler Gamma function, and $\rho_0$ is the central density:
\begin{equation}
    \rho_0 = r_{\rm Pop\RomanNumeralCaps{3}} \times M_{\rm BH}\frac{3-\alpha}{4\pi}\left(\frac{\sigma_{\rm inf}^2}{GM_{\rm BH}}\right)^3
\end{equation}
Here $r_{\rm Pop\RomanNumeralCaps{3}}=0.07$ denotes the mass fraction of Pop \RomanNumeralCaps{3} stars among all stellar populations in a galaxy at $z=10$ \citep{Magg22}. The loss cone filling factor $q$ can also be described as follow:
\begin{equation}
    q(E, m_\star) = \nu(m_\star) \bigg(\frac{E}{\sigma_{\rm inf}^2}\bigg)^{\alpha-4},
\end{equation}
with
\begin{equation}
    \begin{split}
        \nu (m_\star) &= \frac{8\sqrt{\pi}}{3} (3-\alpha) \frac{\gamma(\alpha+1)}{\gamma(\alpha-1/2)} \\
        & \bigg[\frac{5}{32(\alpha-1/2)} + \frac{3I_B(1/2,\alpha) - I_B(3/2, \alpha)}{4\pi} \bigg] \bigg( \frac{G\langle m_{\rm scat}^2\rangle }{\sigma_{\rm inf}^2 \langle m_\star \rangle R_T(m_\star)} \bigg) \ln \Lambda.
    \end{split}
\end{equation}
Here $\ln \Lambda = \ln(0.4 M_{BH}/m_\star)$ is the coulomb logarithm,  $\langle m_\star \rangle = \int{m_\star \Phi_{\rm PopIII}(m_\star) dm_\star}$, 
%is the average stellar mass of the Pop \RomanNumeralCaps{3} stars.  
and $m_{\rm scat}$ represents the total population of stars inside a galaxy that are available for scattering. To simplify the calculation, we assume all Pop III star TDEs are produced by the scattering between a Pop \RomanNumeralCaps{3} star and a Pop I star star, so that 
\begin{equation}
    \langle m_{\rm scat}^2 \rangle = \int m_\star^2\Phi_{\rm norm}(m_\star) dm_\star. \\
\end{equation}
Furthermore, $I_B$ is defined as
\begin{equation}
    I_B (\frac{n}{2}, \alpha) = \int_0^1{t^{-\frac{n+1}{2}}(1-t)^{3-\alpha}B(t,\frac{n}{2},\alpha - \frac{1}{2})dt}
\end{equation}
with $B$ as the incomplete Euler Beta function. 

As discussed in Sec \ref{sec:rate}, the volumetric rate of TDE is very mildly affected by the IMF of Pop \RomanNumeralCaps{3} stars. This is mainly because that total TDE rate scales with $\langle m_{\rm scat}^2 \rangle/\langle m_\star \rangle^2$, which does not differ much when the average is taken over the same mass range ($M_\star=30M_\odot-300M_\odot $) for different IMFs.

\counterwithin{figure}{section}
\section{Different TDE Outflow and Emission Models} \label{app:comparison2}

\begin{figure*}
\begin{center}
     \includegraphics[width=8.5cm]{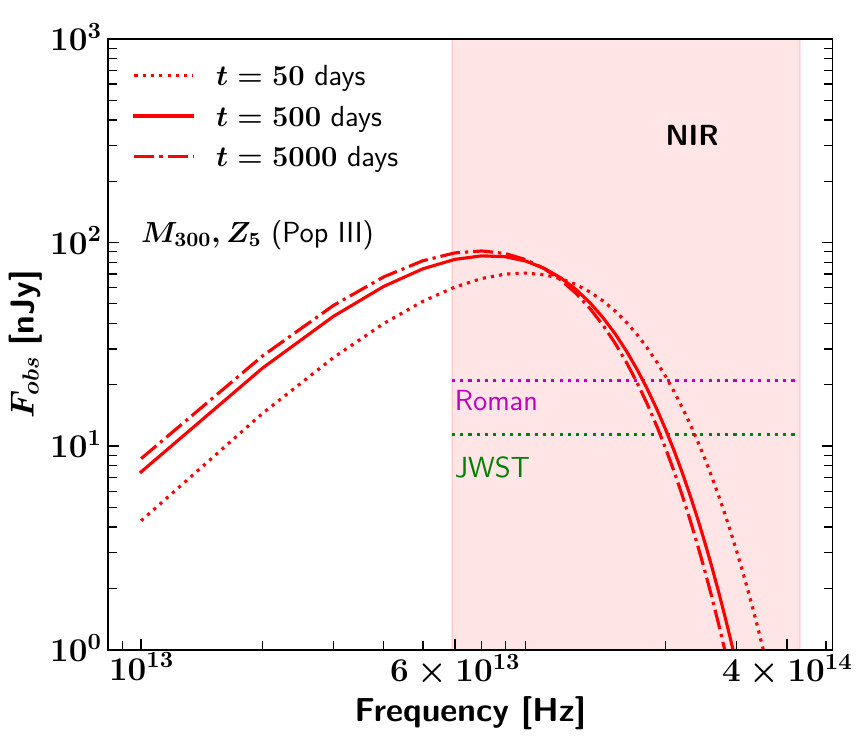}
    \caption{The observed flux of a Pop \RomanNumeralCaps{3} TDE ($M_\star=300M_\odot,~Z=10^{-5},~M_{\rm BH}=10^6M_\odot$) at $z=10$ using the ZEBRA model.}
    \label{fig:obs_zeb}
    \end{center}
\end{figure*}

\noindent In this work, we follow the TDE outflow and emission model proposed by \cite{Strubbe09}. We note that there exist other models for the super-Eddington outflow properties in TDEs. Here we make a comparison with the ZEro-BeRnoulli Accretion (ZEBRA) flow model \citep{CB14}. In particular, we show the observed flux of a $300 M_\odot$ Pop III TDE at $z=10$  based on the ZEBRA model in Fig. \ref{fig:obs_zeb}. Comparing this with Fig. \ref{fig:flux} (top left panel), one finds that both models produce similar observed flux level. However one can note that the SED peaks at slightly higher frequencies and evolves more slowly with time if adopting the ZEBRA model as compared to the SQ09 model.

\listofchanges

\bibliography{pop3}{}
\bibliographystyle{aasjournal}

\end{document}